# Investigating S-type asteroid surfaces through reflectance spectra of Ordinary Chondrites

J. Eschrig[1a], L. Bonal[1], M. Mahlke[2], B. Carry[2], P. Beck[1], J. Gattacceca[3]

[1]Institut de Planétologie et d'Astrophysique de Grenoble, Université Grenoble Alpes, CNRS CNES, 38000 Grenoble (France)

[2]Université Côte d'Azur, Observatoire de la Côte d'Azur, CNRS, Laboratoire Lagrange, France

[3]CNRS, Aix Marseille Univ, IRD, Coll France, CEREGE, Aix-en-Provence, France

Revised version

[a] Corresponding author: jolantha.eschrig@univ-grenoble-alpes.fr (J. Eschrig)
Keywords: Meteorites, Asteroids, surfaces, Spectroscopy, Mineralogy

# Abstract

The search for asteroidal parent bodies of chondrites through various techniques is an ongoing endeavor. A link between ordinary chondrites (OCs) and S-type asteroids has previously been established by the sample return of the Hayabusa space mission. OCs are the class with the most abundant samples in our meteorite collection. We present an in-depth study of the reflectance spectra of 39 equilibrated and 41 unequilibrated ordinary chondrites (EOCs and UOCs). We demonstrate that consistent measuring conditions are vital for the direct comparison of spectral features between chondrites, otherwise hampering any conclusions. We include a comparison with a total of 466 S-type asteroid reflectance spectra from various databases. We analyze (i) if a difference between EOCs and UOCs as well as between H, L and LL can be seen, (ii) if it is possible to identify unequilibrated and equilibrated S-type asteroid surfaces and (iii) if we can further constrain the match between OCs and S-type asteroids all based on reflectance spectra.

As a first step, we checked the classification of the 31 Antarctic UOCs analyzed in the present work, using petrography and magnetic measurements, and evidenced that 74% of them were misclassified. Reflectance spectra were compared between EOCs and UOCs as well as between H, L and LL chondrites using a set of spectral features including band depths and positions, peak reflectance values, spectral slopes and the Ol/(Ol + Px) ratio. UOCs and EOCs reflectance spectra show no clear-cut dichotomy, but a continuum with some EOCs showing stronger absorption bands and peak reflectance values, while others are comparable to UOCs. Moreover, we show by the example of 6 EOCs that their band depths decrease with decreasing grain size. Based on reflectance spectra alone, it is thus highly challenging to objectively identify an unequilibrated from an equilibrated S-type surface. There is no clear distinction of the chemical groups: only LL EOCs of petrographic type > 4 can be distinguished from H and L through less deep 2000 nm band depths and 1000 nm band positions at longer wavelengths. No dichotomy of S-type asteroids can be seen based on the Ol/(Ol + Px) ratio. Their average Ol/(Ol + Px) ratio matches EOCs better than UOCs. A principal component analysis (PCA) was performed illustrating that both the unknown degree of space weathering and the unknown regolith grain size on asteroid surfaces hinder the distinction between equilibrated and unequilibrated surfaces. Lastly, an anti-correlation between the diameter of the asteroids and their 1000 nm band depth is found indicating that larger sized S-type asteroids show finer grained surfaces.

## 1. Introduction

The investigation of asteroids is an important and extremely useful tool for uncovering the formation and evolution history of the Solar System. As many asteroid types have escaped differentiation they are primitive objects (Vernazza et al., 2017). Studying them through laboratory measurements of chondrites, which are readily available on Earth, and remotely through reflectance spectroscopy are common practices. Within these studies, the search for asteroidal parent bodies of chondrites through various techniques is an ongoing endeavor.

Ordinary chondrites (OCs) are the class with the most abundant samples in our meteorite collection ("Meteoritical Bulletin Database," 2021). As such, they have been studied widely, giving us a good understanding of their chemical and petrographic properties. They are sub-divided into equilibrated and unequilibrated ordinary chondrites (EOCs and UOCs, respectively) based on their thermal metamorphic grade. As a measure for the metamorphic grade, chondrites are assigned petrologic types (PT) with UOCs representing PT 3.0-3.9 and EOCs PT 4-7. Moreover, to reflect their mineralogical variability, OCs are further subdivided into H, L and LL based on their total average iron and iron in metal abundance in decreasing order (Krot et al., 2014). The oxidation state increases from H to L to LL.

A link between S-type asteroids and ordinary chondrites has previously been proposed through reflectance spectroscopy (e.g.Binzel et al. (2004); Chapman and Salisbury (1973); Moroz et al. (1996)). This has been confirmed by the Hayabusa space mission that sampled the S-type asteroid Itokawa that is comparable to LL4, LL5 and LL6 type ordinary chondrites (Nakamura et al., 2011).

Besides sample return missions, reflectance spectroscopy is a powerful tool for comparing asteroids with meteorites and looking for links between them (e.g. Sears et al., 2021). Both OCs and S-type asteroids can show variable reflectance spectra. Close to 500 S-type asteroid reflectance spectra are available in the 500-2400 nm wavelength range from various databases (Mahlke et al., in preparation). Equally, 64 and 60 reflectance spectra have previously been obtained for UOCs and EOCs, respectively and made available in the "Reflectance Experiment LABoratory" database ("RELAB," 2021). However, differences in measuring conditions between OC spectra in the RELAB database impair the comparability between samples.

The goal of the present work is to further investigate the link between S-type asteroids and OCs by comparing a large set of S-type asteroid spectra (466) with a large set of OC spectra (80) which have been acquired under consistent measuring conditions. We analyzed the mineralogical variability of the OCs taking into account their post-accretion history (Bonal et al., 2016). With OCs being mineralogically and petrographically diverse, our objective is to investigate whether differences can be seen as well through the tools used for the observation of asteroids. For this a set of spectral features (including band depths and positions as well as spectral slopes) has been determined for each spectrum following previous works (Cloutis et al., 1986; Eschrig et al., 2021). A principal component analysis (PCA) has been performed to investigate the effect of space weathering and regolith grain size on different constituents of the S-type spectra. If a spectral difference between EOCs and UOCs can be found, it raises the question whether a spectral difference between equilibrated and unequilibrated S-type asteroid surfaces can be seen.

## 2. Samples and Methods

### 2.1. Sample list and experimental procedures

This work includes reflectance spectra of 41 UOCs (Eschrig et al., 2019) and of 39 EOCs (Eschrig et al., 2020) (Figs. 1 and 2). Ten out of the 41 UOCs are falls while the remaining 31 are Antarctic finds. For EOCs, 16 are falls and 23 are finds. Samples were provided from different Institutes and Museums as indicated in Tables 1 and 2. Additionally, Tables 1 and 2 show the petrologic type (PT), which was previously determined through Raman spectroscopy on polyaromatic carbonaceous matter (Bonal et al., 2016) for UOCs and the one given in the Meteoritical Bulletin ("Meteoritical Bulletin Database," 2021) for EOCs. To allow for the comparison between meteorites and asteroids, a total of 466 S-type asteroid reflectance spectra (of 323 unique S-type asteroids) were taken from a compilation of reflectance spectra by Mahlke et al., in preparation. The asteroid data in that work was compiled from online resources such as the Small Main-Belt Asteroid Spectroscopic Survey[b] as well as data published in the literature and provided by personal correspondence with the observers. The compilation of asteroid reflectance spectra will be made publicly available upon publication of the Mahlke et al. work. The subset of S-type asteroid spectra used in this work was selected based on the taxonomic classifications of the asteroids found in the literature via the Virtual Observatory Solar System Open Database Network[c] service. We note that the spectral appearance of some asteroids is closer to the olivine-rich A-types than the S-type given in the literature. Some asteroids have more than one spectrum observed. 71 out of these 466 S-type spectra belong to asteroids which have previously been assigned to asteroid dynamical families.

The assessed classification into H, L and LL of each of the UOCs was verified by petrography and magnetic measurements (Table 1). The magnetic susceptibility (MS) was measured at CEREGE (France) on bulk UOCs using an Agico MFK1 instrument operating with a field of 200 A/m and a frequency of 976 Hz. Thin sections and/or thick polished sections (depending on availability) were obtained for each UOC. For petrographic observations, we used a Leica DM2500P petrographic optical microscope at CEREGE. This allowed for the determination of the modal abundance of the different chondrite components (metal, matrix, magnetite, sulfides (troilite) and chondrules/chondrule fragments) through point counting. The degree of terrestrial weathering of each UOC was determined by assessing the amount of metal and troilite weathering products (iron oxides and oxyhydroxides) using the scale defined by Wlotzka (1993). The shock stage of each UOC was determined by looking at thin section in transmitted light and using the shock indicators from Bennett and McSween (1996); Scott et al. (1992) and Stöffler et al. (1991). Mosaic pictures in reflected and/or transmitted light of the sections were used to determine the average chondrule size. Chondrules were manually outlined in the images using a graphics editor. The average chondrule apparent diameter was then determined for each sample. For EOCs, we use the classification given in the Meteoritical Bulletin, since their classification is more straightforward and unquestionable than that of UOCs.

---

[b] https://smass.mit.edu
[c] https://ssp.imcce.fr/webservices/ssodnet/

Reflectance spectra were acquired using the SHADOWS instrument (Potin et al., 2018), a spectro-radio goniometer available at IPAG (Grenoble, France). Measurements were done on powdered chondrites. For this, chondrites were manually crushed into a fine powder and transferred to a sample holder located in an environmental cell. Subsequently, reflectance spectra were measured on the flat, smoothed surface of the powders, under vacuum (P < $10^{-4}$ mbar) and at 80°C temperature. Evacuating and heating the samples reduces the terrestrial water contamination, leaving the hydration feature of the spectra predominantly due to chondritic hydration. Spectra were acquired between 340 – 4200 nm at a 20 nm spectral resolution. The measuring geometry of $i$ = 0° and $e$ = 30° was chosen according to previous works (Eschrig et al., 2021). The reflectance spectra were normalized to the Lambertian surface using a Spectralon$^{TM}$ standard between 340 and 2100 nm and an Infragold$^{TM}$ standard between 2100 and 4200 nm. All spectra were photometrically corrected for reflection effects induced by the sapphire window used to close the environmental cell (Potin et al., 2020).

To evaluate the effect that grain size has on the shape of EOC reflectance spectra, a small series of test was performed in which 6 EOC (Kernouvé (H6), Ste Marguerite (H4), NWA 12961 (L7), Monte das Forte (L5), Moshampa (LL5) and Los Vientos 423 (H6)) and one UOC (RBT 04251 (LL 3.4)) were first ground by hand (as all UOCs and EOCs presented in this work) and then ground at 30 Hz using a laboratory ball mill for 5 min and lastly for 10 min. Sample Kernouvé was additionally ground in the ball mill for 20 min at 30 Hz. At each stage, the grain size was estimated by looking at the powder under an optical binocular and measuring the largest grains visible at 5-10 different spots. A reflectance spectrum was measured for each powder using the measuring conditions described previously.

### 2.2. Analytical procedure

For the classification of the UOCs, the results of MS, chondrule apparent diameter and modal abundances were considered. For the MS measurements, the reference values of $\log(\chi)$ = 5.24±0.08 (n = 10) for H3, $\log(\chi)$ = 4.79±0.11 (n=11) for L3 and $\log(\chi)$ = 4.41±0.16 (n=14) for LL3 were used, with $\chi$ in $10^{-9}$ $m^3kg^{-1}$ (Table 1) (unpublished updated version of Rochette et al., 2003). Since increasing terrestrial weathering results in a decrease of the magnetic susceptibility of ordinary chondrites, the magnetic classification for finds must take their weathering grade into account (Rochette et al., 2012).

The metal modal abundance determined by point counting was compared to the average abundances in H: 7.80±0.62 vol% (n=25), L: 3.54±0.34 vol% (n=53) and LL: 1.16±0.59 vol% (n=14) as computed by Gattacceca et al. (2014) and using meteorite densities from Consolmagno et al. (2008). For the average chondrule apparent diameter (Table 1) the reference values of H: 450 µm, L: 500 µm and LL: 690 µm (Metzler, 2018) were used.

A set of spectral features has been determined to ease the comparison process of reflectance spectra. The reflectance spectra of OCs are characterized by three main bands at 1000, 2000 and 3000 nm. They are due to olivine showing a strong absorption feature around 1000 nm and pyroxene having an absorption feature both around 900 nm and 2000 nm (e.g. Gaffey, 1976). The exact position of the 2000 nm band depends of the relative abundance of low- to high-Ca pyroxene (namely, ortho- and clinopyroxene). Mixtures dominated by low-Ca pyroxene show 2000 nm bands located around 1900 nm while for those dominated by high-Ca pyroxene it is located around 2100 nm (Cloutis and Gaffey, 1991; Singer, 1981). In general, the abundance of pyroxene in ordinary chondrites is dominated by orthopyroxene (e.g. Dunn et al. (2010a)). For hydrated samples, a hydration band is present at 3-micron, whose shape and position depends on the type of hydration (phyllosilicates, oxyhydroxide, molecular water). The determination of the spectral features is done as described in Eschrig et al. (2021). The spectral features include the depth and position of the 1000 and 2000 nm bands which were determined by fitting the bands with a polynomial fit around the absorption minima. Then a linear baseline fit was done to the band areas and the band depth was calculated following Clark (1999). The maxima to either side of the 1000 and 2000 nm band were chosen as boundaries for the linear baseline fit. The peak reflectance at around 700 nm was determined. The visual slope at wavelengths lower than 700 nm was determined by calculating the steepest slope in the 340 nm to 520 nm range and linearly fitting the points around this area. We assess the relative abundance of the end-members olivine (Ol) and pyroxene (Px) in the samples, as previously done e.g. by Vernazza et al. (2014). For this the Ol/(Ol + Px) ratio is determined according to Dunn et al. (2010b) by calculating the Band Area Ratio (BAR) of the 1 and 2 micron band (Band 2 Area over Band 1 area) and then using the following equation:

$$\frac{Ol}{(Ol + Px)} = -0.424 \times BAR + 0.728$$

All spectral values are reported in Tables 3 and 4.

Space Weathering (SW) is a process that affects the asteroid surface through bombardment with micro-meteorites, and irradiation by solar wind and cosmic ions (Brunetto et al., 2006). The effect of SW on the reflectance spectra of silicon-rich S-type asteroids is described as reddening and darkening of the spectra (Marchi et al., 2005) while not significantly changing the position or relative intensities of the mafic silicate bands (Brunetto et al., 2006). It affects only the most upper layer of the asteroid surface (Pieters and Noble, 2016). As the chondrite travels through the Earth's atmosphere, any surface signature of SW is removed. To overcome this difference between asteroids and chondrites, two approaches can be used: i) "de-space weathering" the asteroid spectra analytically or ii) artificially "space weathering" the meteorite spectra. Approach ii) has the disadvantage that the amount of artificial SW that needs to be added to each spectrum is unknown. For approach i) on the other hand, irradiation experiments have shown that L6 chondrites that were space-weathered in the laboratory experienced strong modification of their spectral slope but negligible changes of their 1000 nm and 2000 nm bands (Brunetto et al., 2006; Salisbury and Hunt, 1974). Brunetto et al. (2006) shows that the ratio between the space weathered (sw) and non-space weathered (non-sw) spectrum can be fitted with an exponential function

$$\frac{sw\ spectrum}{non-sw\ spectrum} = K \cdot e^{\frac{C_S}{\lambda}}$$

with $K$ being a scaling factor, $C_S$ being the strength of the SW and $\lambda$ the wavelength. Under the assumptions that SW mainly influences the slope of the spectra (Brunetto et al., 2006; Salisbury and Hunt, 1974) and that the overall slope has no contributions other than SW, the asteroid spectra are thus "de-space weathered" by fitting the spectra with an exponential function and dividing the SW spectrum by this fit. After this correction, the same spectral features as mentioned above were determined for the asteroid spectra.

We further perform a Principal Component Analysis (PCA) of the OC and asteroid spectra (Fig. 8). PCA is a dimensionality reduction technique which projects the input data into a lower-dimensional space along a set of orthogonal vectors that maximize the variability of the projected data points. PCA is particularly suited for exploration and visualization of high-dimensional datasets such as reflectance spectra. In essence, PCA finds the components responsible for the largest variance in the data and allows to visualize them separately. These components are referred to as principal components (PCs) (Hotelling, 1933; Pearson, 1901) and are mathematically given by the eigenvectors of the covariance matrix of the input data. Since the wavelength-resolution of the OC spectra is higher than for the S-type spectra, we compute the PCs based on the latter and project the meteorite spectra into the resulting lower-dimensional space. Computing the PCs based on the OC spectra would lead to PCs that may contain high-frequency signals which the lower-resolution asteroid spectra cannot resolve. After determining the PCs, the principal score of each sample can be calculated. It is given by the vector product of the spectrum and the corresponding PC. PCA was computed on the demeaned spectra, hence the score of zero in all components indicates the position of the mean S-type spectrum. Spectra which are, e.g., redder than the average have a positive principal score. PCA requires a complete data matrix as input. Therefore, we could only compute it based on 339 of the 466 S-type spectra. The remaining ones had to be dropped due to unobserved or later-removed reflectance values at the edges of the wavelength range. We further excluded 7 spectra which are likely misclassified A-types and presented as outliers in the PCA results. As PCA is based on the covariance structure of the data, it is strongly susceptible to outliers.

## 2.3. Importance of consistent measuring conditions

The majority (53) of the UOC spectra that are already available in the RELAB database were measured under consistent measuring conditions by Vernazza et al. (2014). In their work the powdered samples were first sieved ensuring particle sizes between 0 and 45 µm. Then the measurements were performed presumably at ambient temperature and pressure (information not systematically mentioned on the RELAB database).

The samples measured in the present work have not been sieved to avoid introducing a bias in chondrite components. Metal is less easily ground than matrix material resulting in a possible depletion of metal in sifted chondrite powders. Therefore, the grain size, being sub-millimetric in our work as determined during microscope inspection, exceeds that of the samples in Vernazza et al. (2014). Furthermore, our dataset was systematically measured under vacuum and at 80°C to mimic the highly desiccating environment present at the surface of asteroids.

When comparing the UOC spectra that appear in both datasets (Fig. 3) it clearly appears that the spectra measured by Vernazza et al. (2014) show consistently smaller band depths (Fig. 3a) and higher peak reflectance values (Fig. 3b). This is consistent with the fact that reflectance increases and band depths become shallower with decreasing grain size (as shown for instance by Mustard and Hays, 1997 and Sultana et al., 2021 for pure olivine and pyroxene).

For EOCs, the measuring conditions of the reflectance spectra in the RELAB database vary largely, ranging from bulk powder to bulk piece measurements as well as presenting varying measuring temperatures, atmospheres and geometries ("RELAB", 2021). When comparing the spectra of EOCs that are both in the RELAB database and in the dataset presented in the present work there is no clear trend. Indeed, the 1000 nm and 2000 nm band depths of the RELAB data range from less deep to comparable to deeper than those of the dataset in the present work (Fig. 3a). The overall reflectance at 700 nm of the RELAB spectra can range from lower to comparable to higher than those of our dataset as well (Fig. 3b). This corroborates the fact that the grain sizes of the EOC spectra in the RELAB database are either smaller than or comparable to our data set. The difficulty of comparison of EOC spectra from the RELAB database to the present dataset is therefore strongly increased. We conclude that consistent measuring conditions are vital when directly comparing spectral features between chondrites, otherwise hampering any conclusions. In the present work, we thus decided not to include RELAB spectra of OCs.

## 3. Results

### 3.1. Classification of UOCs

The classification of UOCs is not as straightforward as the classification of EOCs. Indeed, the absence of equilibration of olivine and pyroxene in UOCs results in the electron probe microanalyses, which are classically used to separate equilibrated H, L, and LL chondrites, not being readily useable to separate H3, L3, and LL3 chondrites. This especially applies to the most primitive UOCs. The most accurate way of classifying UOCs is, therefore, through petrographic indicators, such as metal abundance (determined by microscopy, or through the measurement of magnetic susceptibility), and chondrule size. For the classification of UOCs from ANSMET (the Antarctic Search for Meteorites program that provided the Antarctic UOCs in the present work), these indicators were either not in common use or no information was given on the procedure of classification at all. We, therefore, checked the classification of all Antarctic UOCs used in the present study.

The results of the magnetic susceptibility and chondrule apparent diameter measurements used to verify the classification of UOCs are listed in Table 1. The abundance of metal and iron oxides (used as a proxy of the initial abundance of metal, neglecting a possible minor contribution from the weathering of troilite), as well as the terrestrial weathering degree determined for each UOC is given in Table 1 as well. Some UOCs were classified solely based on a few of the methods mentioned above as shown in Table 1.

The abundance of metal and iron oxides, as determined by point counting, ranges from 1.2 vol% for Krymka (LL3.2) to 12 vol% for EET 83248 (H>3.6). The average apparent chondrule diameter ranges from 285 µm for MAC 88174 (H>3.6) to 901 µm for Piancaldoli (LL3.1, Marrocchi et al., 2020). log($\chi$) ranges from 3.83 for EET 90066 (LL3.1) to 5.23 for WIS 91627 (H>3.6). Lastly, the weathering grade of the UOCs range from W0 for un-weathered samples to W3 for the most weathered sample DOM 08468 (H3.6). Any differences to the initial classifications of H, L and LL (given in the "Meteoritical Bulletin Database," (2021)) were altered accordingly. As a result, 14 Antarctic UOCs (ALH 83008, ALH 84086, DOM 03287, EET 87735,

EET 90066, EET 90628, GRO 06054, LEW 87248, LEW 87284, LEW 88617, LEW 88632, MET 00489, MIL 05050, MIL 05076) initially classified as L were found to be LL or LL(L), 5 Antarctic UOCs (BTN 00302, LAR 04382, MCY 05218, MET 00506, RBT 04251) initially classified as H were found to be LL or L, 3 Antarctic UOCs (ALH 76004, LAR 06469, ALH 83010) initially classified as LL found to be LL(L) or L and 1 Antarctic UOC (EET 96188) initially classified as L(LL) found to be LL, adding up to a total of 23 miss-classified Antarctic UOCs among the 31 considered. This is equivalent to a miss-classification rate of 74 % for Antactic UOCs. Additionally, the fall Bishunpur, which was initially classified as LL, was found to be L/LL.

### 3.2. EOC and UOC spectra

In Figures 1 and 2 the reflectance spectra of all UOCs and EOCs measured in the present work are shown. For better visibility the spectra were offset along the y-axis and sorted by PT with increasing PT from bottom to top. UOC spectra are plotted in the 500-4200 nm wavelength range while EOCs are plotted in the 500-2600 nm wavelength range. It immediately becomes apparent that on average EOCs (Fig. 2) exhibit stronger absorption bands than UOCs (Fig. 1). UOCs show the so called 1000 nm absorption band at wavelengths shorter than 1000 nm. The 2000 nm absorption band is weak. EOCs have very pronounced 1000 nm absorption bands which are located below 1000 nm for H and shift towards longer wavelengths from H to L to LL (Fig. 2). The exception is NWA 13838 (L5) which has been shock darkened ("Meteoritical Bulletin Database," 2021) and will therefore not be considered in the following discussion. The 2000 nm band of EOCs is broad with varying intensity. The shape and location of the 1000 nm and 2000 nm bands match the presence of olivine and low-Ca pyroxene. As discussed in Section 2.2, spectra dominated by orthopyroxene show 2000 nm band positions around 1900 nm, which is the case for all OCs considered in the present work (Figs. 1 and 2, Tables 3 and 4). The exception is LAR 06469 (L >3.6) for which the position is close to 2100 nm. We, thus, conclude that the pyroxene abundance of all but possibly one of the OCs considered in the present work is dominated by orthopyroxene.

19 of the EOCs show an increase of reflectance at wavelengths shorter than 500 nm. This is an experimental artifact related to the measuring instrument. Because of the steep decrease in the light source intensity below 500 nm, the reflectance system becomes sensitive to background signal below this wavelength, in particular for dark samples. This introduces a strong signal at low wavelengths.

The shock values of UOCs range from S1 to S3 with the exception of BTN 00302 (LL 3.1-3.4) and LAR 06279 (LL 3.05-3.1) which have shock stages of S4/S5 (Table 1). Shock darkened samples are identified through the existence of melt pockets and veins of metal and sulfides in olivine and pyroxene fractures. The shock pressures estimated to be needed for this process corresponds to shock stages between S5 and S6 (Kohout et al., 2020). Some small melt pockets and shock veins were visible in the thin sections of samples BTN 00302 and LAR 06279. Out of these two samples, only BTN 00302 seems to show a rather flat, featureless spectrum (Fig. 1) as is common for shock darkened chondrites.

As mentioned above, EOC NWA 13838 (L5) is shock darkened as stated in the "Meteoritical Bulletin Database" (2021) which is confirmed by the absence of spectral features in Fig. 2. Furthermore, the presence of shock veins and melt pockets is mentioned for EOCs Iran 009 (LL5), Moshampa (LL5), NWA 8275 (LL7) and Viñales (L6) ("Meteoritical Bulletin Database," 2021). None of these samples show noticeable difference in band depths in comparison to the other EOCs considered in the present work (Fig. 2).

Therefore, with the exception of BTN 00302 (LL 3.1-3.4) and NWA 13838 (L5), we conclude that none of the OCs considered in the present work experienced shock darkening. This does not mean that shock did not impact their spectra at all, but no obvious consequence of shock can be observed.

### 3.2.1. Reflectance spectral features

In Tables 3 and 4, the reflectance spectral features determined from the spectra as explained in Section 2.2 are given for UOCs and EOCs, respectively. Listed are the visual slope at wavelength lower than 700 nm, the 700 nm peak reflectance value, the 1000 nm and 2000 nm band depths and band positions and the Ol/(Ol + Px) ratio. The visual slopes range from 1.4E-4 nm$^{-1}$ (Chainpur (LL3.4)) to 6.44E-4 nm$^{-1}$ (LAR 06279 (LL 3.05)) for UOCs and from 7.59E-5 nm$^{-1}$ (NWA 12475 (LL6)) to 9.32E-4 nm$^{-1}$ (Iran 009 (LL5)) for EOCs. The 700nm peak reflectance ranges from 9.4 % (EET 87735 (LL3.05)) to 22.5 % (Parnallee (LL3.6)) for UOCs and from 9.4 % (NWA 12475 (LL6)) to 45.1 % (Monte Das Forte (L5)) for EOCs. The 1000 nm band is located between 920 nm and 1020 nm for UOCs and EOCs. The depth of the 1000 nm band ranges from 9.0 % (MIL 05050 (LL3.1)) to 20.7 % (Parnallee (LL3.6)) for UOCs and from 7.91 % (NWA 8268 (L4)) to 46.95 % (Kernouvé (H6)) for EOCs. The 2000 nm band is located between 1840 nm and 2020 nm for UOCs and between 1870 nm and 1970 nm for EOCs. The exception is UOC LAR 06469 (L>3.6) which shows a 2000 nm band position at 2098 nm. The 2000 nm band depths range from 3.5 % (Krymka (LL3.2)) to 10.4 % (ALH 84086 (LL3.8)) for UOCs and from 3.52 % (NWA 8628 (L4)) to 21.24 % (Moshampa (LL5)) for EOCs. Lastly, the Ol/(Ol + Px) ratio ranges from 18.8 % (Bremervörde (H(L)3.9)) to 54.3 % (LEW 88617 (LL3.6)) for UOCs and from 10.8 % (Coya Sur 001 (H5)) to 65.3 % (Saint-Séverin (LL6)) for EOCs.

### 3.2.2. Grain size effects on reflectance spectral features

The approximate grain size of the powders of EOCs Kernouvé (H6), Ste Marguerite (H4), NWA 12961 (L7), Monte das Forte (L5), Moshampa (LL5) and Los Vientos 423 (H6) and UOC RBT 04251 (LL3.4) for different grinding times are listed in Table 5. For EOCs, the grain size of the powders decreases from being hand ground to being ground for 5 min in the ball mill to being ground for 10 min (20 min) in the ball mill. After hand-grinding, the powders are coarse grained when examining them under an optical binocular. After 5 min in the ball mill, the powder becomes significantly finer. Overall, it consists of a fine grayish/whitish powder (~a few µm grain size) mixed with several dark grains that can have much larger sizes (see Table 5). In some cases, the fine-grained powder initially sticks together forming a fine grained "sand cake"-like structure but can be broke apart easily using a needle. After 10 min in the ball mill, the powders become darker to the naked eye. They show a similar texture to the powder ground for 5 min but less large dark grains can be seen mixed with the fine-grained grayish/whitish powder. The maximum size of the largest grains decreases even more (see Table 5). The same can be said after grinding Kernouvé for 20 min. For the UOC RBT 04251 (LL 3.4), on the other hand, the grain size after hand grinding is already small enough, that it is comparable to the grain size of EOCs after 5-10 min of grinding in the ball mill (Table 5). After 5 minutes of grinding the UOC in the ball mill the grain size decreases and becomes <150 µm. Figure 4a shows the reflectance spectra of the OCs at different grain sizes. The spectral features determined from the spectra are listed in Table 5. As can be seen, spectral features immediately become much smaller after 5 min of grinding in the ball mill. Especially for UOC RBT 04251 the spectrum practically becomes featureless. Therefore, the UOC is not ground for longer periods of time. For EOCs, basically no features are left in the spectra after 10 min or even 20 min (Kernouvé) in the ball mill (Figure 4a). The 1000 nm band depths decrease by 91.41 % of their original depth for Kernouvé (H6) after 20 min of grinding and by 86.46 % of their original depth for NWA 12961 (L7) after 10 min. For the 2000 nm band depths similar decreases can be observed. The 1000 nm band positions tend to shift to shorter wavelengths with decreasing grain size while the 2000 nm band position seems to decrease in some cases and remain comparable in other cases (Table 5). It needs to be mentioned here that the determination of the band position becomes less accurate as the band depths decrease since the signal to noise ratio (SNR) becomes more important in comparison to the spectral features. Interestingly, the 700 nm peak reflectance decreases with grain size as well, dropping to more than half of the original value for most samples after 10 min of grinding. The visual slope becomes less steep with decreasing grain size (Table 5). Due to the SNR and the increase of reflectance at wavelengths shorter than 500 nm due to an experimental artifact (see Section 3.2) the visual slope could not be determined for samples ground longer than 5 min.

### 3.2.3. S-type Asteroid reflectance spectral features

The reflectance spectra of the 466 S-type asteroids which were included in the present work as a comparison to OCs are shown in Figure 5. Shown are both the raw and de-space weathered S-type asteroid spectra. As expected (Brunetto et al., 2006), de-space weathering the spectra mainly modifies the spectral slope. Looking at the de-space weathered S-type spectra, a low spectral variability can be seen (Fig. 5b). The visual slope ranges from $1.11E-4$ $nm^{-1}$ to $2.35E-3$ $nm^{-1}$ and the 700 nm peak relative reflectance from 99.4 % to 112.9 %. The 1000 nm and 2000 nm band depths vary from 7.21 % to 41.57 % and 1.08 % to 20.51 %, respectively. The 1000 nm and 2000 nm band positions are located between 870 nm to 1070 nm and between 1630 nm to 2220 nm, respectively. Lastly, the Ol/(Ol + Px) ratio ranges from 0.46 % to 70.72 %. All spectral values related to S-type asteroids are available online at the CDS[d].

## 4. Discussion

### 4.1. Comparing EOC and UOC reflectance spectra

EOCs are thermally metamorphosed UOCs. With increasing metamorphic grade, the chondrites progressively approach chemical equilibrium, transferring elements such as iron from the matrix to chondrule silicates (e.g., Krot et al. (2014)). There is, indeed, a decrease of the total abundance of metal (mostly present in the matrix) as well as a loss of small grained metal particles in the matrix. This modification of the iron valence state (from $Fe^0$ to $Fe^{2+}$ and $Fe^{3+}$) influences the position and depth of the 1000 nm and 2000 nm absorption bands. The mafic silicate bands become deeper and shift to shorter wavelengths with increasing metamorphic grade. The equilibration process starts at lower temperatures for olivine than for pyroxene (Scott and Jones, 1990) because of a higher diffusivity of Fe-Mg in olivine. Thus, it is expected to see an increase of band depth and a small shift to shorter wavelengths (less than 50 nm) for the 1000 nm olivine band even with less metamorphosed OCs.

---

[d] Data is only available at the CDS via anonymous ftp to cdsarc.u-strasbg.fr (130.79.128.5) or via http://cdsarc.u-strasbg.fr/viz-bin/cat/J/XXX

As expected, the 1000 nm and 2000 nm bands of the EOCs are deeper than those of UOCs with band depths increasing with increasing petrologic type (PT) (see Fig 6 a and b, p-values of 2.48E-10 and 3.09E10-6, respectively). There are some exceptions (12 EOCs out of 39) to this trend especially for the 2000 nm band. Specifically, samples Coya Sur 001 (H5), Catalina 024 (H4), Catalina 309 (L5), El Medano 378 (L6), Los Vientos 083 (H4), Los Vientos 147 (L6), Los Vientos 155 (H5), Los Vientos 416 (L4), Los Vientos 432 (H5), NWA 8628 (L4), Paposo 012 (H6) and Tamdakht (H5) show very similar 1000 nm and 2000 nm band depths to UOCs. Samples Tuxtuac (LL5), Bensour (LL6), Bandong (LL6), Saint-Severin (LL6), NWA 7283 (LL6), NWA 12475 (LL6), NWA 8275 (LL7) and NWA 12546 (LL7) show stronger 1000 nm band depths but 2000 nm band depths similar to UOCs. As expected, the increase in band depth is seen more clearly for the 1000 nm band than for the 2000 nm band in EOCs. Overall, the band positions are comparable between UOCs and EOCs (Fig. 6c and d, p-values of 0.433 and 0.0319, respectively). The 1000 nm band position of a few LL EOCs exceeds those of the UOCs. In the 2000 nm region, EOCs collectively show band positions between 1900 nm and 1970 nm. The 2000 nm band positions of UOCs ranges from 1840 nm to 2020 nm making their band positions comparable or lower than those of EOCs. The outlier, LAR 06469 (L>3.6) at 2098 nm shows a secondary band around 2400 nm shifting the 2000 nm band position to longer wavelengths. The peak reflectance at around 700 nm is on average higher for EOCs than for UOCs (Fig. 6e, p-values of 1.26E-9). Lastly, the visual slope at wavelengths lower than 700 nm of EOCs is comparable to slightly steeper than for UOCs (Fig. 6f, p-values of 3.5E-3).

The comparison of EOC and UOC reflectance spectra, therefore, shows that while there is some spectral variability, there seems to be more of a continuum of spectral signatures. EOCs can show deeper 1000 nm and 2000 nm band depths and higher peak reflectance values but several EOC spectra are also comparable to UOC spectra. In addition, the band positions are comparable between UOCs and EOCs. Overall, the maximum 2D KS p-value of these sample pairs was 0.433, indicating a statistical difference in the underlying distributions.

### 4.2. Comparing H, L and LL reflectance spectra

The reflectance spectra of OCs are dominated by contributions of the iron-bearing mineral phases olivine (Ol) and pyroxene (Px) (Vernazza et al., 2014 and ref. therein) in the 1000 nm and 2000 nm region. The Ol/(Ol + Px) ratio is thus particularly relevant for spectral comparisons between OCs. In particular Vernazza et al. (2014) found that the Ol/(Ol + low-Ca Px) ratio of H decreases significantly in comparison to those of L and LL with increasing metamorphic grade. This observation allowed them to distinct H from L and LL and to advocate for the existence of two S-type asteroid groups: H-like and LL-like (see more in Section 4.3). In the following section we explore the spectral differences we could observe between H, L and LL.

Among EOCs, LL are spectrally distinct from H and L. Indeed, when considering the 1000 nm and 2000 nm band depths, EOC LLs Tuxtuac (LL5), Bensour (LL6), Bandong (LL6), Saint-Severin (LL6), NWA 7283 (LL6), NWA 12475 (LL6), NWA 8275 (LL7) and NWA 12546 (LL7) plot separately (Fig. 7a). Additionally, EOC LL show 1000 nm band positions around 980-1020 nm while for H and L the position is shifted to shorter wavelengths around 920-980 nm (Fig. 6c). The exceptions to this rule are the LL Iran 009 (LL5), Moshampa (LL5), NWA 12556 (LL5) and Soko Banja (LL4), having a 1000 nm band positions around 960-970 nm, making them comparable to H and L. While H and L of PT>5 show signs of deepening of both the 1000 nm and 2000 nm band as well as shifting of the 1000 nm band to shorter wavelengths due to the equilibration process, LL seem to be less affected by this process and only show a deepening of the 1000 nm band in comparison to UOCs. Since LL type have a lower total metal abundance than H and L (Scott and Krot, 2003), they are less effected by the decrease of the total iron in metal along the metamorphic sequence and hence show a smaller shift of the 1000 nm band to shorter wavelengths and less deep 2000 nm band depths in comparison to H and L type EOCs. For UOCs no clear spectral separation between H, L and LL can be seen.

In Figure 7b, the Ol/(Ol + Px) ratios of the EOCs and UOCs are plotted over their PT. The majority of OCs between PT3-4 have Ol/(Ol + Px) ratios between 0.3 and 0.6. A few UOCs show lower ratios <0.3. In contrast to the work by Vernazza et al. (2014) no systematic difference in the Ol/(Ol + Px) ratio between H and L, LL is observed. For PT 5-7 an increase in ratios can be seen up to 0.7. It seems that within each PT group ≥ 4, H show lower ratios than L and LL with the exception of Moshampa (LL5), Limon Verde 004 (L6) and El Médano 378 (L6) which show low Ol/(Ol + Px) ratios within their group despite being L and LL, not H. For UOCs no distinction of H, L and LL based on the Ol/(Ol + Px) ratio is possible.

The different conclusions in the present work and in Vernazza et al. (2014) can be explained by the use of different classifications and PT values. The classification into H, L and LL of the UOCs in the present work has been reassessed through magnetic susceptibility and petrography: more than half (24/41) of the UOCs were reclassified in the present work (see Table 1 and Section 3.1). Out of the 16 UOCs which were considered both in the present work and by Vernazza et al., (2014), 10 have been reclassified. Moreover, the PT values of the UOCs in this work were taken from Bonal et al. (2016) based on the structural order of the polyaromatic carbonaceous matter assessed by Raman spectroscopy. They showed in their study that the PT initially assigned for the sample declaration in the Meteoritical Bulletin are not necessarily correct as most of the time they only reflect preliminary descriptions of samples. Lastly, with the reclassification of 24 out of the 41 UOCs we only have 8 H, 4 L but 29 LL or LL(L) UOCs in the present work. This disproportionate representation of the different classes among UOCs increases the difficulty of finding systematic differences between them.

In conclusion, LL of PT ≥ 5 can be distinguished from H and L based on their 2000 nm band depth and 1000 nm band position which is shallower and shifted to longer wavelengths for LL in comparison to H and L. This is in agreement with previous works which found that LL can be recognized based on their Ol/(Ol + Px) ratio (Dunn et al., 2010b). The Ol/(Ol + Px) ratios of EOCs can slightly exceed those of UOCs for PT>4. However, the majority of the spectral indicators used here are comparable between EOCs and UOCs. This is in contrast to the observations made by Vernazza et al. (2014) according to which the ratio becomes lower with increasing metamorphic grade for H. A slight separation between H, L and LL is visible within each PT groups ≥ 4 of EOCs for our data, with ratios decreasing from LL and L to H.

### 4.3. Ordinary chondrites versus S-type asteroids

The link between S-type Asteroids and EOCs has previously been confirmed by the sample return of the Hayabusa-1 space mission (Nakamura et al., 2011). In their work on UOCs, Vernazza et al. (2014) found that S-type asteroids can be sub-divided into two distinct compositional groups: H-like and LL-like S-type asteroids. This conclusion was based on the observation of a clear separation of the Ol/(Ol + Px) ratio of H versus L and LL UOCs and a bimodality in the Ol/(Ol + Px) ratios of a large set (83) of S-type asteroid spectra.

In the present work we have found that while EOCs can have distinct spectral features from UOCs there seems to be more of a continuum, with the exception of the 2000 nm band position which is comparable between all UOCs and EOCs (Section 4.1). In this section we explore if the observations of Vernazza et al. (2014) can be reproduced with our larger dataset. Will the dichotomy between H and L, LL UOCs be visible for EOCs as well? Lastly, can a difference between equilibrated and unequilibrated S-type asteroid surfaces be seen?

#### 4.3.1. PCA and space weathering

For the comparison of meteorite to asteroid reflectance spectra, space weathering plays an important role. As explained in Section 2.2, space weathering was here removed from the S-type asteroid spectra before determining the spectral values. However, the unknown degree of space weathering on the asteroids surfaces yields a systematic uncertainty on the spectral parameters which cannot be quantified without detailed knowledge of the asteroids surface properties. In the following, the influence of space weathering is thus analyzed as a gradual change of the asteroid spectral properties via PCA.

The first three PCs of the raw S-type asteroid spectra are shown in Figure 8a. The percentages assigned to each component in the legend state the explained variance (EV) of the component. The EV quantifies the ratio of variance retained in the projection of the data along the PC to the total variance in the data. As can be seen, the PC1 contributes to about 78 % of the whole information in the asteroid spectra, while PC2 and PC3 contribute 11.9 % and 5.3 %, respectively.

PC1 resembles the overall slope of the asteroid spectra with a 1000nm olivine band. Comparing the PC1 to the raw asteroid spectra in Figure 5, it becomes apparent that the main variability within S-type spectra is given by their slope. PC2 and PC3 in Figure 8a resemble the absorption features of olivine and orthopyroxene, respectively. Disregarding the slope, the main difference between the 466 S-type spectra is given by their surface mineralogy, which is to be expected.

Figures 8b, c and d show the principal scores of the S-type spectra in comparison to the H, L and LL EOC and UOC spectra. Figure 8b shows the principal scores 1 versus 2. As the PCs have been computed from the raw S-type spectra, the asteroids are centered around the score 0. Most OCs have a negative score 1 as they are less red than the average raw S-type. The positive score 2 of most OC indicates an increase in the olivine-like absorption signature around 1000 nm in comparison to S-type spectra. When comparing the principal scores 2 and 3 in Figure 8d the difference in the mineralogical signatures are highlighted: EOCs show much stronger principal score 2 and 3 than UOCs and S-type spectra, indicating a stronger olivine- and orthopyroxene-like signature in EOCs. Furthermore, LL EOCs have considerably stronger principal score 2 (olivine-like) setting them apart from L and H. This is in accordance with the observation based on spectral parameters (Section 4.2).

To quantify the influence of space weathering on the reflectance spectra, we computed the principal scores of the mean S-type spectrum after de-weathering it. The line between the de-weathered mean spectrum and the raw mean (by definition at score 0) is shown as solid black arrow in the principal score plots of Figure 8. We interpret this as the direction in which S-type spectra are shifted with increasing degree of space weathering within each plot. What can be seen, is that de-space weathering the S-type asteroid spectra results in (i) spectral blueing (principal score 1 shifting to negative values) (ii) an increase of olivine-like absorption features (principal score 2 shifting to positive values) and (iii) an increase in orthopyroxene-like features (principal score 3 shifting to positive values). It also shows that a better match between S-type asteroids and OCs can be achieved, depending on how strongly the S-type spectra are de-space weathered or in other words: the more the S-type asteroid gets space weathered, the more it will shift from having spectral features comparable to UOCs, to comparable to EOCs, to not matching at all.

In conclusion, it is important to note that a match between S-type and OC spectra is very dependent on the treatment of the SW in the S-type asteroid spectra. SW mostly influences the slope of S-type spectra. Since such a big variability in slopes is seen for S-type spectra in Figure 5a, in the following parts we will compare OCs with de-space weathered S-type spectra in order to be less sensitive to slope and avoid formulating a hypothesis based on the extent of SW experienced by the asteroid surfaces.

### 4.3.2. Comparison of OCs with de-space weathered S-type asteroids

In Figure 9a the 1000 nm and 2000 nm band depths of the UOC and EOC spectra are compared with those of the de-space weathered S-type asteroid spectra. As can be seen, the majority of the S-type asteroid spectra cluster closer to UOCs and EOCs with faint spectral features. This is consistent with previous works (Eschrig et al., 2021) where the end member S-type spectra taken from DeMeo et al. (2009) match UOCs. In Figure 9b the 700 nm peak reflectance (for spectra normalized to 550 nm) over the 1000 nm band depth of the S-type asteroids is compared to those of UOCs and EOCs. It shows that the majority of UOCs and EOCs show stronger 700 nm peak reflectance values than S-type asteroids.

In Figure 10 a histogram of the Ol/(Ol + Px) ratios of the de-space weathered S-type asteroids considered in the present work is shown. A large peak centered around 50-60% can be seen. No bimodality is visible as was previously observed by Vernazza et al. (2014). The analytical method for determining the Ol/(Ol + Px) ratio used in the present work differs from Vernazza et al. (2014). Here, we used the Band Area Ratio (BAR) to calculate the ratio as indicated in Dunn et al. (2010b). The advantages of this method are that it is simple yet sufficient for the present work. It does not require any assumptions about e.g. grain size since it can be determined directly from the spectra. In contrast, Vernazza et al. (2014) applied a more sophisticated Radiative transfer model to determine the Ol/(Ol + low-Ca Px) ratio. Hypothesis such as the grain size distribution are thus necessary. Furthermore, Vernazza et al. (2014) excluded objects of collisional families in their histogram.

Figure 10 also includes the Ol/(Ol + Px) ratios of all 41 UOCs and 39 EOCs. Two things can be seen: i) UOCs show ratios slightly to the left of the center of the peak of S-type asteroid ratios and ii) EOCs, whose ratios are slightly higher than those of UOCs, fall right in the center of the S-type asteroid ratio peak. This indicates that the average Ol/(Ol + Px) ratio of S-type asteroids match those of EOCs better than those of UOCs.

### 4.3.3. Grain size effects

Interestingly, no S-type matches can be found for the EOCs with strong absorption features based on the 1000 nm and the 2000 nm band depths (Fig 9a). This is consistent with the observations based on PCA (Fig. 8, Section 4.3.1): no match for these EOCs can be found with either the raw or de-space weathered S-type spectra. A possible explanation for this could be a difference in grain size between the powders measured in the laboratory and the actual regolith on the S-type asteroid surfaces. UOCs usually contain more fine-grained and porous matrix material than EOCs and are thus more easily ground into fine powders by hand as shown by the example of RBT 04251 (LL 3.4) (Table 5 and Section 3.2.2).

In Figure 9, the spectral features of 6 EOCs and 1 UOC measured at different grain sizes (see Section 3.2.2) were added into the plots comparing the 2000 nm and 1000 nm band depths (Fig. 9a). They were also added to the 700 nm peak reflectance over the 1000 nm band depth plot in Fig. 9b. As can be seen, decreasing the grain size of the powders moves the 1000 nm and 2000 nm band depths from out of the reach of any S-type asteroid (top right corner) to being comparable to UOC band depths. After 20 min of grinding, the 1000 nm and 2000 nm band depths decrease to even lower values than for UOCs (Fig. 9a). The 700nm peak reflectance decreases with decreasing grain size as indicated by the example of Los Vientos 423 (H6) in Fig. 9b. After 10 min of grinding in the ball mill, the 700 nm peak reflectance of Los Vientos 423 goes from significantly higher than UOC and S-type spectra, to comparable to UOCs.

We further add an indicator of the effect of increasing grain size to the principal score plots in Figure 8. For this the mean of the scores determined from fine grained powders that were ground for 5 min, 10 min and 20 min was taken. The dashed arrows in Figure 8 connects the mean score of the fine-grained EOC spectra to the mean score of the EOCs before additional grinding. We expect the OC spectra to shift in this direction with increasing grain size. It is apparent, in all three score plots, that decreasing the grain size of the EOCs moves them closer to the asteroid distribution. These results highlight the already well-known importance of grain size for the study of asteroid surfaces as well as for the comparison of asteroid and meteorite spectra (e.g., Krämer Ruggiu et al., 2021).

When plotting the diameter of the asteroids considered in the present work as a function of the 1000 nm band depth (Fig. 4b) there seems to be a loose anti-correlation ($r^2$ = 0.215), for asteroids of d > 20 km. This means that larger asteroids tend to show finer grained surfaces, consistent with Delbo et al. (2015) ( and references therein). Therefore, the analysis of grain size effects on spectral features done in the present work appears to be a meaningful explanation of the differences in the observed distributions between asteroids and OCs, and between UOCs and EOCs.

### 4.4. Consequences for S-type asteroids

We have shown so far that i) the spectral differences between UOCs and EOCs are not a clear-cut dichotomy. EOCs can show stronger band depths and higher peak reflectance values than UOCs but depending on their grain size they can also be comparable to UOC spectra (Fig. 6 and 9); ii) the average Ol/(Ol + Px) ratios of S-type asteroids match those of EOCs better than those of UOCs (Fig. 10) and iii) an anti-correlation between the 1000 nm band depth of S-type asteroids and their diameter could be seen (Fig. 4b). These results allow us to make some hypothesis of the structure of S-type asteroids.

#### 4.4.1. Indications on the S-type asteroid structure

Based on the identification of metamorphic sequence among OCs, the common view of S-type asteroid structure is an onion shell model. In the case of an onion shell model, the surface and outer crust of the undisturbed S-type asteroid is expected to be UOC-like, while the main bulk of the material in the center of the asteroid is EOC-like (McSween et al., 2002). In this scenario, the S-type asteroids start off as large, intact bodies and then become brecciated into smaller pieces through impact processes. The result is a few objects which retain the pristine UOC-like surface and a larger number of bodies which show EOC-like surfaces exposed by the brecciation of their previously parent bodies (e.g. McSween Jr. and Patchen, 1989).

In the present work we find that the average Ol/(Ol + Px) ratio of S-type asteroids matches those of EOCs better than those of UOCs (Fig. 10). While it does not definitely prove it, this result is aligned with the presence of a few large asteroids with UOC-like surfaces and many small asteroids with EOC-like surfaces as we would expect for an onion shell model. This hypothesis would also be in line with UOCs being strongly under-represented in our meteorite collection with only about 5.4 % of all OC being UOCs and the remaining being EOCs (based on falls in the "Meteoritical Bulletin Database," (2021)).

## 5. Conclusion

With OCs being mineralogically and petrographically diverse, our objective was to understand whether or not such a diversity is visible among asteroid surfaces based on reflectance spectroscopy. In the present work we conducted an in-depth study of a large number of UOCs using varying measuring techniques to make sure of their classification. It appears that 74 % (23 of the 31) of the Antarctic UOCs were misclassified. The most common misclassification is LL3 chondrites being classified as L3 chondrites. We attribute this high number of mis-classifications to the lack of robust classification data used in the "Meteoritical Bulletin Database," (2021).

Reflectance spectra of 80 OCs (39 EOCs and 41 UOCs) were measured under consistent measuring conditions. The spectral features of UOCs and EOCs were compared and found to not be a clean-cut dichotomy but rather a continuum. Among EOCs, LL of PT ≥ 5 can be distinguished from H and L based on their 2000 nm band depth and 1000 nm band position which is shallower and shifted to longer wavelengths for LL in comparison to H and L. Comparing the Ol/(Ol + Px) ratios of EOCs and UOCs showed that EOCs can slightly exceed those of UOCs for PT>4. However, many of the ratios are comparable between EOCs and UOCs. A slight single separation between H, L and LL is visible within each PT groups ≥ 4, with ratios decreasing from LL and L to H.

We studied the influence of grain size on the reflectance spectra of 6 EOCs and 1 UOC and showed that decreasing the grain size of the powders leads to a significant decrease in 1000 nm and 2000 nm band depth. It also leads to a decrease in the 700 nm peak reflectance and a shift of the 1000 nm band position to shorter wavelengths, possibly due to the comminution of opaque components into fine grains. Overall, it leads to EOC spectra becoming more comparable to UOC spectra. UOCs are ground into fine powders easier by hand than EOCs since they contain a larger quantity of porous matrix material.

Surfaces of S-type asteroids are subject to space weathering which induces a change in slope in their reflectance spectra. Considering this as well as the influence grain size has on reflectance spectra, the comparison between S-type asteroid spectra and OC spectra is difficult.

A large set (466) of de-space weathered S-type asteroid reflectance spectra was analyzed in the same way as the OCs. The comparison with the OC spectra showed that the 1000 nm and 2000 nm band depths of S-type asteroids cluster between EOC and UOCs. The exact match is strongly dependent on the amount of SW removed from the S-type asteroid spectra as illustrated by PCA. We also found that the 1000 nm band depth of S-type asteroids is anti-correlated with their diameter. Since we expect a reduced band contrast with smaller grain size (see Section 4.3.3) this advocates for large bodies having a fine-grained regolith, while smaller bodies have a coarser grained regolith.

All these results are in line with, but do not definitely prove an onion shell model for S-type asteroids. The surface and outer crust of the undisturbed S-type asteroids are expected to be UOC-like, while the main bulk of the material in the center of the asteroid is EOC-like. The majority of the bulk OC material would stem from young dynamical families, where EOC-like bodies are the major contributors to the meteorite production. This could explain the strong under-representation of UOCs among OCs and is supported by the good match between S-type asteroid and EOC Ol/(Ol + Px) ratios.

There are still several open questions surrounding OCs. In the overall representation of H to L to LL in our OC collection only 11.1 % of OCs are LL type (in comparison to 47.6 % being L and 41.2 % being H) (percentages based on number of falls in the "Meteoritical Bulletin Database," (2021)). Misclassification can be a partial but not sole explanation. As discussed in Vernazza et al. (2014) possible other reasons for LL-type samples to be less common may be related to the location and age of the source region as well as the size and type of the body. Furthermore, a pronounced hydration band could be observed for most UOCs in this work (Fig. 1), indicating the presence of hydrated minerals. It would, therefore, be interesting to obtain more S-type asteroid spectra in the 2500-4200 nm range (e.g. during the SPHEREx space mission (Ivezić et al., 2022)), which would allow for the 3-micron band to be used as a further constrain on the S-type-OC match.

## Acknowledgements

This work has been funded by the Centre National d'Etudes Spatiales (CNES-France) and by the ERC grant SOLARYS ERC-CoG2017-771691.

US Antarctic meteorite samples are recovered by the Antarctic Search for Meteorites (ANSMET) program which has been funded by NSF and NASA, and characterized and curated by the Department of Mineral Sciences of the Smithsonian Institution and Astromaterials Acquisition and Curation Office at NASA Johnson Space Center. We thank the Natural History Museum, Vienna, Department of Mineralogy and Petrography for providing us with a thick section of Mezö-Madaras (Inv-Nr. NHMV-N2140) and bulk samples of Mezö-Madaras (NHMV_ID_#3993_B2), Parnallee (NHMV_ID_#6207_B2) and Tieschitz (NHMV_ID_#7152_A).

We also thank the Museum D'Histoire Naturelle National (MNHN), the Arizona State University (ASU), the Centre de Recherches Pétrographiques et Geochimiques (CRPG) at Nancy and lastly the Centre Européen de Recherche et D'Enseignement des Géosciences de l'Environnement (CEREGE), Aix-En-Provence for providing us with the remaining EOC and UOC samples.

# Tables and Figures

*Table 1: List of all UOCs measured in the present work. Shown are the results from point counting (metal + iron oxide abundance in vol%) thin section measurements (average apparent chondrule diameter ± standard deviation (s.d.) in µm) and magnetic susceptibility (MS) measurements (log(x) with x in $10^{-9}$ m$^3$/kg) as well as the weathering grade for each sample. The classification that were attributed based on these results as well as the classification given in the Meteoritical Bulletin are listed. Lastly, the petrographic type (PT) as determined by Bonal et al 2016 (for Piancaldoli Marrocchi et al., 2020) and the providers of the bulk samples measured in this work are mentioned. Samples marked with a (\*) are falls, the others are finds. The superscript "sv" stands for "shock veins" and marks samples for which shock veins and melt pockets were observed.*

| sample | Metal + Iron oxides (vol%) | Average apparent diam. ± s.d. (µm) | MS log(x) | Weathering | Shock stage | Classification (MetBul) | Final classification | PT (Bonal et al. 2016) | Provider bulk samples |
|---|---|---|---|---|---|---|---|---|---|
| **ALH 76004** | 3.1 | 580 ± 274 (n = 54) | 4.52 | W0 | S3/S2+ | LL | **L(LL)** | 3.1–3.4 | NASA |
| **ALH 78119** | 1.9 | 546 ± 316 (n = 39) | 4.43 | W1 | S1 | LL | **LL(L)** | 3.5 | NASA |
| **ALH 83008** | 8.8 | - | 4.35 | W0 | S3 | L | **LL** | >3.6 | NASA |
| **ALH 84086** | 4.8 | 779 ± 464 (n = 26) | 4.43 | W0 | S3 | L | **LL** | 3.8 | NASA |

| Sample | | | | | | | | | |
|---|---|---|---|---|---|---|---|---|---|
| ALH 84120 | 5.5 | 596 ± 371 (n = 51) | 4.64 | W0 | S1 | L | **L** | 3.8 | NASA |
| Bishunpur* | 3.6 | - | 4.67 | W0 | S2 | LL | **L/LL** | 3.2 | ASU |
| Bremervörde* | 5.8 | - | 4.98 | W0 | S2 | H(L) | **H/L** | 3.9 | ASU |
| BTN 00302sv | 2.1 | 670 ± 316 (n = 64) | 4.58 | W0 | S4/S5 | H | **LL** | 3.1–3.4 | NASA |
| Chainpur* | 1.8 | - | 4.46 | W1 | S1 | LL | **LL** | 3.4 | ASU |
| Dhajala* | 10.9 | - | 5.12 | W0 | S1 | H | **H** | 3.8 | ASU |
| DOM 03287 | 2.8 | 791 ± 270 (n = 25) | 4.48 | W1 | S3 | L | **LL** | 3.6 | NASA |
| DOM 08468 | 7.8 | 473 ± 200 (n = 86) | 4.47 | W3 | S1 | H | **H** | 3.6 | NASA |
| EET 83248 | 7.7 | 385 ± 308 (n = 40) | 4.87 | W2 | S1 | H | **H** | >3.6 | NASA |
| EET 87735 | 6.9 | 587 ± 284 (n = 19) | 4.00 | W1 | S1 | L | **LL** | 3.05–3.1 | NASA |
| EET 90066 | - | - | 3.83 | W1 | S2 | LL | **LL** | 3.1 | NASA |
| EET 90628 | - | - | 4.27 | W1 | S3- | LL | **LL** | 3.0 | NASA |
| EET 96188 | - | - | 4.31 | W0 | S3 | L/LL | **LL** | 3.1–3.4 | NASA |
| GRO 06054 | 3.3 | 639 ± 451 (n = 42) | 4.36 | W1 | S2 | L | **LL** | 3.1 | NASA |
| Hallingeberg* | 3.3 | - | 4.78 | W0 | S1/S2 | L | **L** | 3.4 | ASU |
| Kymka* | 1.2 | - | 4.03 | W1 | S3 | LL | **LL** | 3.2 | MNHN, Paris |
| LAR 04382 | 1.6 | 692 ± 324 (n = 72) | 4.23 | W0 | S3 | H | **LL** | 3.1–3.4 | NASA |
| LAR 06279sv | 1.1 | 886 ± 506 (n = 29) | 4.00 | W1 | S4/S5 | LL | **LL** | 3.05–3.1 | NASA |
| LAR 06469 | 10.8 | 456 ± 239 (n = 75) | 4.20 | W2 | S2/S3 | LL | **L** | >3.6 | NASA |
| LEW 87248 | 3.6 | 486 ± 211 (n = 23) | 4.50 | W0 | S3 | L | **L(LL)** | 3.0 | NASA |
| LEW 87284 | - | 684 ± 364 (n = 42) | 4.44 | W0 | S3+ | L | **LL** | 3.1–3.4 | NASA |
| LEW 88617 | 1.1 | 536 ± 312 (n = 24) | 3.86 | W1 | S3 | L | **LL** | 3.6 | NASA |
| LEW 88632 | 1.6 | 788 ± 449 (n = 26) | 4.13 | W1 | S2 | L | **LL** | 3.4 | NASA |
| MAC 88174 | 7.8 | 337 ± 170 (n = 177) | 5.10 | W1 | S3 | H | **H** | >3.6 | NASA |
| MCY 05218 | 6.2 | 614 ± 272 (n = 79) | 4.55 | W2 | S3 | H | **L** | 3.05–3.1 | NASA |
| MET 00489 | - | - | 3.92 | W1 | S2/3 | L | **LL** | 3.05–3.1 | NASA |
| MET 00506 | 6.6 | 581 ± 244 (n = 75) | 4.07 | W1 | S2+/S3 | H | **LL** | 3.1 | NASA |
| Mezö-Madaras* | 3.7 | 590 ± 331 (n = 91) | - | W0 | S4 | L | **L** | 3.7 | NHM, Vienna |
| MIL 05050 | 3.8 | 563 ± 266 (n = 40) | 4.21 | W1 | S3 | L | **LL** | 3.1 | NASA |
| **MIL 05076** | 5.3 | 808 ± 466 (n = 53) | 3.92 | W1 | S3 | L | **LL** | 3.4 | NASA |
| **Parnallee*** | 1.7 | - | 4.49 | W0 | S3 | LL | **LL** | 3.7 | NHM, Vienna |
| **Piancaldoli*** | 2.7 | 901 ± 445 (n = 352) | - | W0 | S2/S3 | LL | **LL** | 3.1 | CRPG, Nancy |
| **RBT 04251** | 3.9 | 746 ± 332 (n = 16) | 4.34 | W1 | S3 | H | **LL** | 3.4 | NASA |
| **Tieschitz*** | 3.7 | 446 ± 289 (n = 23) | - | W0 | S1/S2 | H/L | **H/L** | 3.6 | NHM, Vienna |
| **TIL 82408** | - | - | 4.03 | W1 | S3 | LL | **LL** | 3.05–3.1 | NASA |
| **WIS 91627** | 9.6 | 329 ± 269 (n = 72) | 5.23 | W0 | S2 | H | **H** | >3.6 | NASA |
| **WSG 95300** | 11.2 | 426 ± 247 (n = 41) | 4.89 | W1 | S1 | H | **H** | 3.4 | NASA |

*Table 2: List of all EOCs measured in the present work. Listed are the classification and petrologic type (PT) as given in the Meteoritical Bulletin. Lastly, the providers of the bulk EOCs are mentioned. Samples marked with a (\*) are falls, the others are finds. The superscription "sv" stands for "shock veins" and marks samples for which shock veins and melt pockets were observed. The superscription "sd" stands for "shock darkened".*

| sample | Classification | Providers |
| --- | --- | --- |
| Bandong* | LL6 | MNHN, Paris |
| Beni M'hira* | L6 | CEREGE, Aix-en-Provence |
| Bensour* | LL6 | CEREGE, Aix-en-Provence |
| Catalina 024 | H4 | CEREGE, Aix-en-Provence |
| Catalina 309 | L5 | CEREGE, Aix-en-Provence |
| Coya Sur 001 | H6 | CEREGE, Aix-en-Provence |
| El Medano 378 | L6 | CEREGE, Aix-en-Provence |
| Forest Vale* | H4 | MNHN, Paris |
| Iran 009$^{sv}$ | LL5 | CEREGE, Aix-en-Provence |
| Iran 011 | L6 | CEREGE, Aix-en-Provence |
| Kernouve* | H6 | MNHN, Paris |
| Lançon* | H6 | MNHN, Paris |
| Limon Verde 004 | L6 | CEREGE, Aix-en-Provence |
| Los Vientos 083 | H4 | CEREGE, Aix-en-Provence |
| Los Vientos 147 | L6 | CEREGE, Aix-en-Provence |
| Los Vientos 155 | H5 | CEREGE, Aix-en-Provence |
| Los Vientos 416 | L4 | CEREGE, Aix-en-Provence |
| Los Vientos 423 | H6 | CEREGE, Aix-en-Provence |
| Los Vientos 432 | H5 | CEREGE, Aix-en-Provence |
| Monte Das Forte* | L5 | MNHN, Paris |
| Moshampa*$^{sv}$ | LL5 | CEREGE, Aix-en-Provence |
| Mount Tazerzait* | L5 | MNHN, Paris |
| NWA 12475 | LL6 | CEREGE, Aix-en-Provence |
| NWA 12546 | LL7 | CEREGE, Aix-en-Provence |
| NWA 12556 | LL5 | CEREGE, Aix-en-Provence |
| NWA 12961 | L7 | CEREGE, Aix-en-Provence |
| NWA 13838$^{sd}$ | L5 | CEREGE, Aix-en-Provence |
| NWA 7283 | LL6 | CEREGE, Aix-en-Provence |
| NWA 8628 | L4 | CEREGE, Aix-en-Provence |
| NWA 8275$^{sv}$ | LL7 | CEREGE, Aix-en-Provence |
| NWA 8477 | L5 | CEREGE, Aix-en-Provence |
| Paposo 012 | H6 | CEREGE, Aix-en-Provence |
| Saint-Séverin* | LL6 | MNHN, Paris |
| Soko-Banja* | LL4 | MNHN, Paris |
| Ste. Marguerite* | H4 | MNHN, Paris |
| Tamdakht* | H5 | CEREGE, Aix-en-Provence |
| Taqtaq-e Rasoul* | H5 | CEREGE, Aix-en-Provence |
| Tuxtuac* | LL5 | MNHN, Paris |
| Viñales*$^{sv}$ | L6 | CEREGE, Aix-en-Provence |

*Table 3: List of the spectral features determined for each of the UOCs considered in the present work. Given are the visual slope at wavelengths lower than 700 nm, the 700 nm peak reflectance, the 1000 nm and 2000 nm band depth and positions as well as the Ol/(Ol + Px) ratio. All spectral features where determined as explained in Section 2.2 following Eschrig et al. (2021).*

| sample | Visual slope ($10^{-5}$ nm$^{-1}$) | 700nm Peak refl. (%) | 1000nm Band Depth (%) | 2000nm Band Depth (%) | 1000nm Band Pos. (nm) | 2000nm Band Pos. (nm) | Ol/(Ol+ Px) (%) |
|---|---|---|---|---|---|---|---|
| ALH 76004 | 30.01 ± 0.88 | 13.89 ± 0.01 | 15.64 ± 0.15 | 7.46 ± 0.04 | 960 | 1937 | 38.2 |
| ALH 78119 | 38.99 ± 1.9 | 14.53 ± 0.02 | 16.54 ± 0.49 | 7.87 ± 0.06 | 950 | 1996 | 31.9 |
| ALH 83008 | 39.5 ± 1.85 | 12.62 ± 0 | 12.31 ± 0.12 | 4.34 ± 0.07 | 959 | 1889 | 48.4 |
| ALH 84086 | 36.39 ± 0.81 | 20.07 ± 0.02 | 19.6 ± 0.86 | 10.36 ± 0.09 | 964 | 2005 | 30.4 |
| ALH 84120 | 32.72 ± 0.91 | 20.37 ± 0.03 | 14.01 ± 0.76 | 6.77 ± 0.02 | 942 | 1956 | 30.0 |
| Bishunpur | 20.38 ± 0.39 | 13.03 ± 0.01 | 14.5 ± 0.29 | 4.88 ± 0.05 | 950 | 1890 | 48.0 |
| Bremervörde | 38.16 ± 1.59 | 16.9 ± 0 | 19.07 ± 0.37 | 9.57 ± 0.01 | 928 | 1901 | 18.8 |
| BTN 00302 | 14.35 ± 0.69 | 14.13 ± 0.01 | 11.56 ± 0.28 | 4.78 ± 0.1 | 962 | 1918 | 44.0 |
| Chainpur | 13.97 ± 2.12 | 13.02 ± 0.01 | 13.63 ± 0.08 | 4.81 ± 0.04 | 961 | 1892 | 47.9 |
| Dhajala | 23.16 ± 0.6 | 18.52 ± 0.01 | 19.63 ± 0.73 | 8.19 ± 0.07 | 937 | 1933 | 40.9 |
| DOM 03287 | 35.81 ± 1.36 | 12.97 ± 0 | 13.06 ± 0.2 | 5.46 ± 0.08 | 945 | 1965 | 37.8 |
| DOM 08468 | 26.08 ± 0.85 | 11.41 ± 0 | 12.8 ± 0.37 | 4.75 ± 0.15 | 969 | 1973 | 51.7 |
| EET 83248 | 35.79 ± 1.96 | 11.74 ± 0 | 11.87 ± 0.13 | 6.01 ± 0.08 | 960 | 1945 | 30.4 |
| EET 87735 | 25.93 ± 1.38 | 9.43 ± 0 | 12.72 ± 0.18 | 5.28 ± 0.09 | 960 | 1949 | 37.2 |
| EET 90066 | 33.08 ± 2.57 | 10.34 ± 0 | 11.15 ± 0.08 | 4.84 ± 0.12 | 965 | 1901 | 38.2 |
| EET 90628 | 38.74 ± 1.41 | 13.45 ± 0.01 | 15.01 ± 0.15 | 7.09 ± 0.33 | 964 | 1893 | 40.1 |
| EET 96188 | 53.52 ± 2.8 | 16.24 ± 0 | 14.72 ± 0.25 | 5.59 ± 0.16 | 952 | 1875 | 42.4 |
| GRO 06054 | 34.61 ± 1.47 | 11.74 ± 0.01 | 12.61 ± 0.04 | 4.96 ± 0.08 | 988 | 1916 | 46.5 |
| Hallingeberg | 56.47 ± 3.44 | 18.49 ± 0.01 | 11.59 ± 0.56 | 4.25 ± 0.07 | 953 | 1914 | 39.0 |
| Kymka | 38.54 ± 1.37 | 13.57 ± 0 | 12.35 ± 0.14 | 3.47 ± 0.09 | 961 | 1869 | 54.1 |
| LAR 04382 | 43.08 ± 1.5 | 14.84 ± 0.01 | 17.86 ± 0.59 | 8.29 ± 0.09 | 970 | 1997 | 39.9 |
| LAR 06279 | 64.35 ± 2.75 | 17.65 ± 0.02 | 15.07 ± 0.7 | 6.92 ± 0.11 | 938 | 1896 | 36.3 |
| LAR 06469 | 33.76 ± 1.6 | 13.19 ± 0 | 11.66 ± 0.09 | 5.3 ± 0.03 | 964 | 2098 | 31.7 |
| LEW 87248 | 26.12 ± 0.87 | 14.21 ± 0 | 11.4 ± 0.15 | 4.66 ± 0.05 | 950 | 1867 | 39.6 |
| LEW 87284 | 31.72 ± 1.01 | 15.48 ± 0 | 12.34 ± 0.32 | 4.45 ± 0.1 | 927 | 1841 | 51.3 |
| LEW 88617 | 38.85 ± 1.24 | 13.39 ± 0 | 14.34 ± 0.21 | 4.73 ± 0.06 | 962 | 1969 | 54.3 |
| LEW 88632 | 48.24 ± 2.79 | 14.22 ± 0.02 | 16.01 ± 0.47 | 5.13 ± 0.19 | 980 | 2008 | 49.8 |
| MAC 88174 | 31.66 ± 0.69 | 13.96 ± 0.02 | 13.33 ± 0.29 | 5.34 ± 0.07 | 962 | 1976 | 41.3 |
| MCY 05218 | 41.34 ± 1.34 | 12.37 ± 0 | 10.53 ± 0.27 | 3.58 ± 0.08 | 948 | 1861 | 47.1 |
| MET 00489 | 43.28 ± 1.26 | 17.34 ± 0 | 19.01 ± 0.27 | 8.46 ± 0.16 | 968 | 1946 | 42.0 |
| MET 00506 | 38.31 ± 2.05 | 11.93 ± 0 | 14.23 ± 0.1 | 4.98 ± 0.08 | 980 | 2019 | 39.3 |
| Mezö-Madaras | 34.68 ± 0.67 | 17.22 ± 0.01 | 13.89 ± 0.43 | 4.65 ± 0.1 | 940 | 1858 | 53.2 |
| MIL 05050 | 32.34 ± 1.3 | 11.37 ± 0 | 9.02 ± 0.6 | 3.93 ± 0.65 | 952 | 1889 | 48.0 |
| MIL 05076 | 30.69 ± 1.34 | 10.74 ± 0 | 11.81 ± 1.97 | 3.83 ± 0.2 | 940 | 1913 | 52.3 |
| Parnallee | 59.1 ± 7.25 | 22.56 ± 0.02 | 20.71 ± 3.01 | 7.33 ± 0.23 | 958 | 1929 | 47.8 |
| Piancaldoli | 26.14 ± 0.37 | 15.84 ± 0 | 9.43 ± 0.06 | 3.65 ± 0.13 | 949 | 1944 | 41.9 |
| RBT 04251 | 49.63 ± 1.95 | 16.29 ± 0 | 13.65 ± 0.21 | 5.57 ± 0.05 | 958 | 1907 | 38.3 |
| Tieschitz | 29.92 ± 2.59 | 16.29 ± 0 | 14.55 ± 0.12 | 5.44 ± 0.04 | 979 | 1981 | 46.2 |
| **TIL 82408** | 33.91 ± 1.18 | 12.52 ± 0 | 12.01 ± 0.08 | 5.19 ± 0.03 | 955 | 1894 | 36.4 |
| **WIS 91627** | 51.92 ± 3.25 | 15.61 ± 0 | 15.54 ± 0.6 | 7.7 ± 0.08 | 937 | 1887 | 28.2 |
| **WSG 95300** | 31.74 ± 1.08 | 11.17 ± 0.01 | 11.48 ± 0.15 | 3.76 ± 0.06 | 977 | 1951 | 53.2 |

Table 4: List of the spectral features determined for each of the EOCs considered in the present work. Given are the visual slope at wavelengths lower than 700 nm, the 700 nm peak reflectance, the 1000 nm and 2000 nm band depth and positions as well as the Ol/(Ol + Px) ratio. All spectral features where determined as explained in Section 2.2 following Eschrig et al. (2021).

| sample | Visual slope ($10^{-5}$ nm$^{-1}$) | 700nm Peak refl. (%) | 1000nm Band Depth (%) | 2000nm Band Depth (%) | 1000nm Band Pos. (nm) | 2000nm Band Pos. (nm) | Ol/(Ol+ Px) (%) |
|---|---|---|---|---|---|---|---|
| Bandong | 47.09 ± 1.27 | 25.38 ± 0.01 | 33.43 ± 0.13 | 6.92 ± 0.39 | 1000 | 1946 | 64.6 |
| Beni M'hira | 72.72 ± 2.14 | 31.32 ± 0.04 | 33.48 ± 3.13 | 11.18 ± 1.42 | 952 | 1923 | 56.3 |
| Bensour | 29.63 ± 2.48 | 25.14 ± 0.01 | 32.04 ± 0.59 | 7.01 ± 0.46 | 994 | 1943 | 63.7 |
| Catalina 024 | 54.29 ± 4.27 | 18.06 ± 0.01 | 18.43 ± 0.89 | 7.49 ± 0.33 | 943 | 1905 | 38.7 |
| Catalina 309 | 37.28 ± 4.35 | 19.18 ± 0.02 | 20.88 ± 0.24 | 6.38 ± 0.16 | 970 | 1956 | 55.7 |
| Coya Sur 001 | 27.62 ± 2.65 | 16.85 ± 0.02 | 9.44 ± 0.33 | 5.18 ± 0.04 | 942 | 1957 | 10.9 |
| El Medano 378 | 18.44 ± 2.57 | 15.99 ± 0.02 | 13.93 ± 0.18 | 5.39 ± 0.05 | 965 | 1949 | 45.6 |
| Forest Vale | 26.36 ± 0.61 | 21.27 ± 0 | 22.89 ± 2.53 | 12.61 ± 1.68 | 931 | 1908 | 31.0 |
| Iran 009 | 93.19 ± 7.57 | 25.33 ± 0.02 | 27.91 ± 0.62 | 10.42 ± 0.29 | 968 | 1949 | 47.2 |
| Iran 011 | 45.8 ± 4.19 | 27.88 ± 0.02 | 26.92 ± 0.31 | 9.19 ± 0.72 | 954 | 1942 | 54.8 |
| Kernouve | 71.82 ± 1.5 | 29.61 ± 0.01 | 46.95 ± 3.35 | 20.8 ± 0.44 | 945 | 1923 | 45.6 |
| Lançon | 32.89 ± 1.52 | 37.62 ± 0.04 | 30.8 ± 8.46 | 14.86 ± 4.06 | 938 | 1910 | 42.6 |
| Limon Verde 004 | 55.26 ± 4.04 | 22.21 ± 0.03 | 32.06 ± 1.69 | 14.54 ± 0.15 | 951 | 1929 | 40.8 |
| Los Vientos 083 | 40.34 ± 4.91 | 16.36 ± 0.03 | 13.46 ± 0.04 | 6.38 ± 0.32 | 924 | 1872 | 35.0 |
| Los Vientos 147 | 62.3 ± 3.19 | 20.46 ± 0.02 | 18.31 ± 0.58 | 6.16 ± 0.46 | 959 | 1912 | 52.6 |
| Los Vientos 155 | 50.19 ± 4.22 | 18.43 ± 0.02 | 16.35 ± 0.08 | 7.69 ± 0.29 | 929 | 1896 | 34.4 |
| Los Vientos 416 | 29.7 ± 2.84 | 17.4 ± 0.02 | 14.27 ± 0.72 | 6.16 ± 0.26 | 960 | 1927 | 47.5 |
| Los Vientos 423 | 77.14 ± 5.82 | 23.9 ± 0.02 | 24.18 ± 2.27 | 9.61 ± 0.4 | 945 | 1898 | 45.0 |
| Los Vientos 432 | 56.35 ± 2.88 | 20.37 ± 0.02 | 19.33 ± 1.5 | 9.05 ± 0.51 | 937 | 1917 | 31.6 |
| Monte Das Forte | 58.56 ± 2.62 | 45.12 ± 0.01 | 38.34 ± 8.01 | 18.87 ± 1.06 | 947 | 1930 | 42.8 |
| Moshampa | 49.06 ± 1.66 | 34.39 ± 0.01 | 39.88 ± 3.58 | 21.24 ± 0.93 | 957 | 1943 | 39.1 |
| Mount Tazerzait | 50.11 ± 1.27 | 30.58 ± 0.01 | 37.67 ± 1.76 | 13.11 ± 0.23 | 963 | 1943 | 54.3 |
| NWA 12475 | 7.59 ± 1.16 | 9.38 ± 0.04 | 38.6 ± 0.23 | 8.51 ± 0.13 | 988 | 1955 | 63.1 |
| NWA 12546 | 27.18 ± 0.52 | 16.05 ± 0 | 20.42 ± 0.14 | 5.68 ± 0.08 | 988 | 1949 | 59.6 |
| NWA 12556 | 47.68 ± 8.26 | 32.44 ± 0.03 | 28.05 ± 3.01 | 11.23 ± 2 | 971 | 1961 | 49.2 |
| NWA 12961 | 80.77 ± 2.17 | 35.53 ± 0.01 | 45.48 ± 1.68 | 16.3 ± 0.53 | 977 | 1936 | 54.5 |

| sample | | Visual slope (10⁻⁵ nm⁻¹) | 700nm Peak refl. (%) | 1000nm Band Depth (%) | 2000nm Band Depth (%) | 1000nm Band Pos. (nm) | 2000nm Band Pos. (nm) | Ol/(Ol+Px) (%) |
|---|---|---|---|---|---|---|---|---|
| NWA 13838 | | -0.01 ± 1.11 | 10.18 ± 0.04 | 9.24 ± 0.46 | 6.32 ± 0.02 | 939 | 1964 | 6.7 |
| NWA 7283 | | 26.46 ± 4.75 | 27.39 ± 0.04 | 36.31 ± 0.21 | 8.96 ± 0.2 | 995 | 1966 | 61.7 |
| NWA 8628 | | 11.56 ± 5.97 | 11.95 ± 0.02 | 7.91 ± 0.46 | 3.52 ± 0.19 | 947 | 1912 | 36.3 |
| NWA 8275 | | 43.74 ± 1.99 | 26.72 ± 0.01 | 29.92 ± 0.44 | 5.11 ± 0.42 | 1012 | 1968 | 64.9 |
| NWA 8477 | | 52.66 ± 3.31 | 33.54 ± 0.02 | 30.75 ± 3.14 | 13.68 ± 0.41 | 953 | 1949 | 43.6 |
| Paposo 012 | | 68.72 ± 4.24 | 21.93 ± 0.02 | 20.5 ± 0.91 | 8.33 ± 0.33 | 945 | 1911 | 38.7 |
| Saint-Séverin | | 10.21 ± 0.61 | 17.57 ± 0 | 25.92 ± 0.35 | 5.03 ± 0.3 | 994 | 1954 | 65.3 |
| Soko-Banja | | 40.43 ± 1.84 | 29.73 ± 0.01 | 31.02 ± 3.26 | 14.75 ± 0.75 | 972 | 1964 | 45.1 |
| Ste. Marguerite | | 27.73 ± 0.65 | 21.77 ± 0 | 25.56 ± 1.58 | 15.27 ± 0.35 | 941 | 1946 | 20.8 |
| Tamdakht | | 40.44 ± 0.83 | 24.53 ± 0 | 19.16 ± 2.85 | 10.12 ± 1 | 944 | 1919 | 39.6 |
| Taqtaq-e Rasoul | | 24.56 ± 1.19 | 32.15 ± 0 | 31.63 ± 5.85 | 15.1 ± 0.72 | 939 | 1916 | 38.2 |
| Tuxtuac | | 78.11 ± 1.01 | 33.82 ± 0.21 | 38.52 ± 0.68 | 11.7 ± 0.35 | 1001 | 1961 | 59.0 |
| Viñales | | 69.55 ± 2.09 | 29.76 ± 0.04 | 42.16 ± 1.96 | 14.83 ± 0.33 | 957 | 1942 | 53.5 |

*Table 5: List of the spectral features determined for EOCs at different grain sizes. Listed are samples measured after being hand ground (hg), after being ground for 5 min in a ball mill, after being ground for 10 min in a ball mill and after being ground for 20 min in a ball mill. Given are the visual slope at wavelengths lower than 700 nm (for spectra where it could be determined), the 700 nm peak reflectance, the 1000 nm and 2000 nm band depth and positions as well as the Ol/(Ol + Px) ratio. All spectral features where determined as explained in Section 2.2 following Eschrig et al. (2021)*

| sample | approx. size of biggest grains (μm) | Visual slope (10⁻⁵ nm⁻¹) | 700nm Peak refl. (%) | 1000nm Band Depth (%) | 2000nm Band Depth (%) | 1000nm Band Pos. (nm) | 2000nm Band Pos. (nm) | Ol/(Ol+Px) (%) |
|---|---|---|---|---|---|---|---|---|
| Kernouve hg | <1000 | 66.51 ± 2.94 | 29.61 ± 0.01 | 46.99 ± 3.6 | 20.84 ± 0.44 | 949 | 1924 | 45.4 |
| Kernouve 10min | <500 | - | 22.29 ± 0.01 | 8.57 ± 0.9 | 3.71 ± 0.53 | 932 | 1894 | 46.3 |
| Kernouve 20min | <150 | - | 12.23 ± 0.01 | 4.03 ± 0.04 | 2.38 ± 0.01 | 944 | 1962 | 21.4 |
| Monte Das Forte hg | <1000 | 65.14 ± 5.61 | 45.12 ± 0.01 | 38.55 ± 0.03 | 19.89 ± 0.11 | 956 | 1934 | 42.8 |
| Monte Das Forte 5min | <500 | 10.73 ± 1.38 | 37.2 ± 0.01 | 16.85 ± 21.81 | 9.13 ± 2.26 | 930 | 1946 | 22.7 |
| Monte Das Forte 10min | <200 | - | 25.93 ± 0.01 | 9.73 ± 0.53 | 4.77 ± 0.29 | 929 | 1933 | 24.7 |
| Moshampa hg | <800 | 49.06 ± 1.66 | 34.39 ± 0.01 | 39.88 ± 3.58 | 21.24 ± 0.93 | 957 | 1943 | 39.1 |
| Moshampa 5min | <500 | 39.68 ± 5.26 | 21.49 ± 0.01 | 15.71 ± 0.54 | 9.29 ± 0.22 | 926 | 1982 | -4.2 |
| Moshampa 10min | <400 | - | 14.79 ± 0.01 | 6.84 ± 0.42 | 3.54 ± 0.2 | 928 | 1947 | 20.4 |
| NWA 12961 hg | <200 0 | 80.77 ± 2.17 | 35.53 ± 0.01 | 45.48 ± 1.68 | 16.3 ± 0.53 | 977 | 1936 | 54.5 |
| NWA 12961 5min | <150 0 | 8.01 ± 1.35 | 26.6 ± 0.03 | 11.13 ± 3.73 | 3.01 ± 0.48 | 946 | 1917 | 54.6 |
| NWA 12961 10min | <350 | - | 15.17 ± 0.01 | 6.16 ± 0.04 | 1.99 ± 0.28 | 933 | 1919 | 51.9 |
| Ste Marguerite hg | <104 0 | 27.73 ± 0.65 | 21.77 ± 0 | 25.56 ± 1.58 | 15.27 ± 0.35 | 941 | 1946 | 20.8 |
| Ste Marguerite 5min | <700 | 3.02 ± 0.51 | 19.2 ± 0.01 | 9.02 ± 3.02 | 3.56 ± 0.17 | 931 | 1931 | 37.0 |
| Ste Marguerite 10min | <500 | - | 13.55 ± 0.01 | 4.67 ± 0.03 | 3.28 ± 0.04 | 914 | 1889 | -27.2 |
| Los Vientos 423 hg | <100 0 | 77.14 ± 5.82 | 23.9 ± 0.02 | 24.18 ± 2.27 | 9.61 ± 0.4 | 945 | 1898 | 45.0 |
| Los Vientos 423 5 min | <500 | 62.85 ± 3.32 | 22.25 ± 0.01 | 10.2 ± 0.89 | 4.51 ± 0.06 | 931 | 1973 | 12.8 |
| Los Vientos 423 10 min | <400 | 39.17 ± 3.35 | 15.4 ± 0.01 | 5.48 ± 0.15 | 3.54 ± 0.04 | 948 | 2059 | 4.1 |
| RBT 04251 hg | <400 | 49.63 ± 1.95 | 16.29 ± 0 | 13.65 ± 0.21 | 5.57 ± 0.05 | 958 | 1907 | 38.3 |
| RBT 04251 5 min | <150 | 15.91 ± 2.12 | 14.08 ± 0.01 | 3.72 ± 0.05 | 1.43 ± 0.06 | 929 | 2081 | 37.6 |

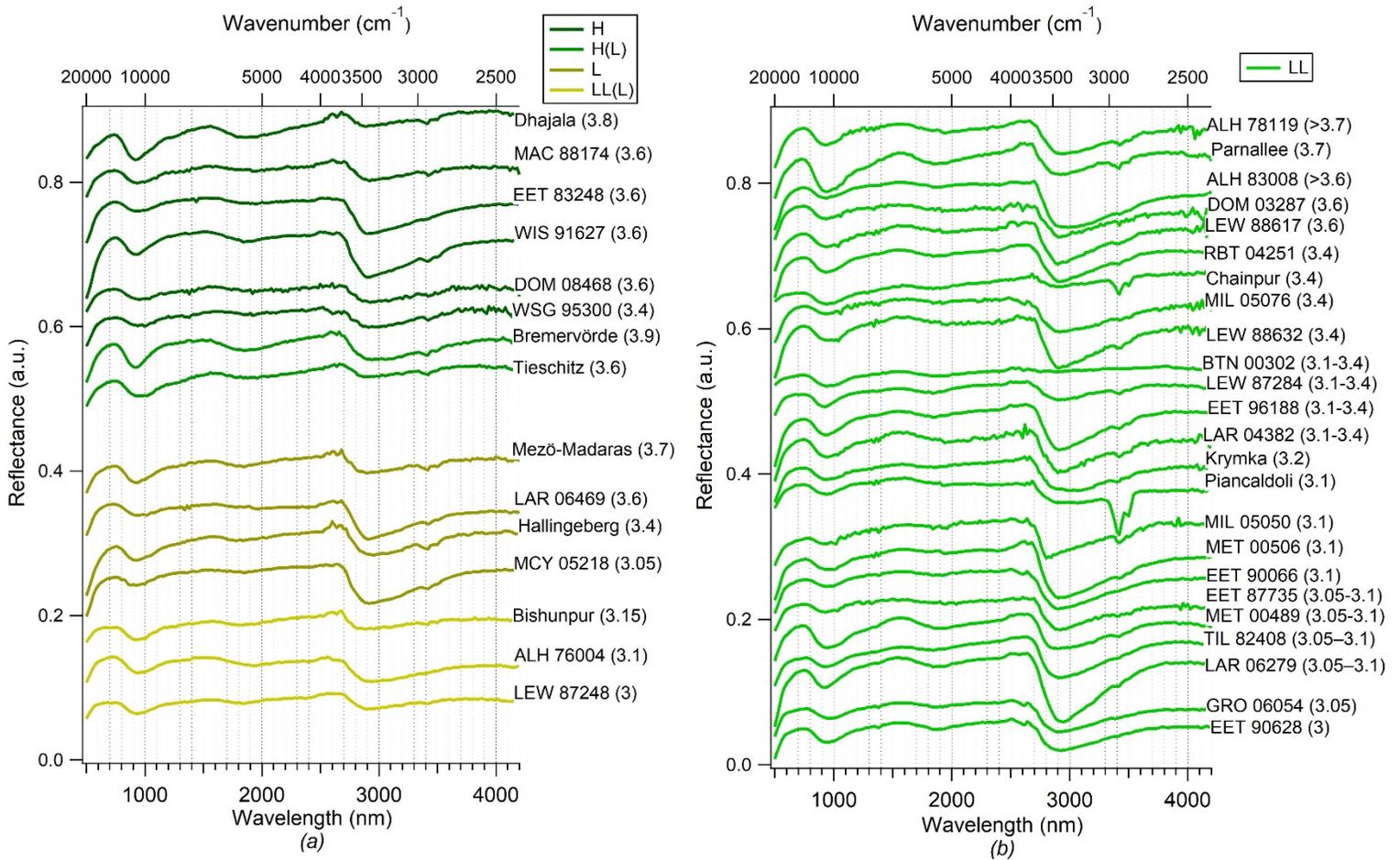

*Figure 1: Reflectance spectra of the 41 UOCs measured in the present work in the 500 – 4200 nm wavelength range. The spectra are shifted along the y-axis for better visibility and sorted by metamorphic grade within each class (H, H(L), L, LL(L) and LL) with increasing metamorphic grade from bottom to top as indicated by the Petrographic type (PT) values (Bonal et al., .2016) given in parenthesis behind each sample name. The classification used in this graph is the same as determined in the present work (see Table 1)*

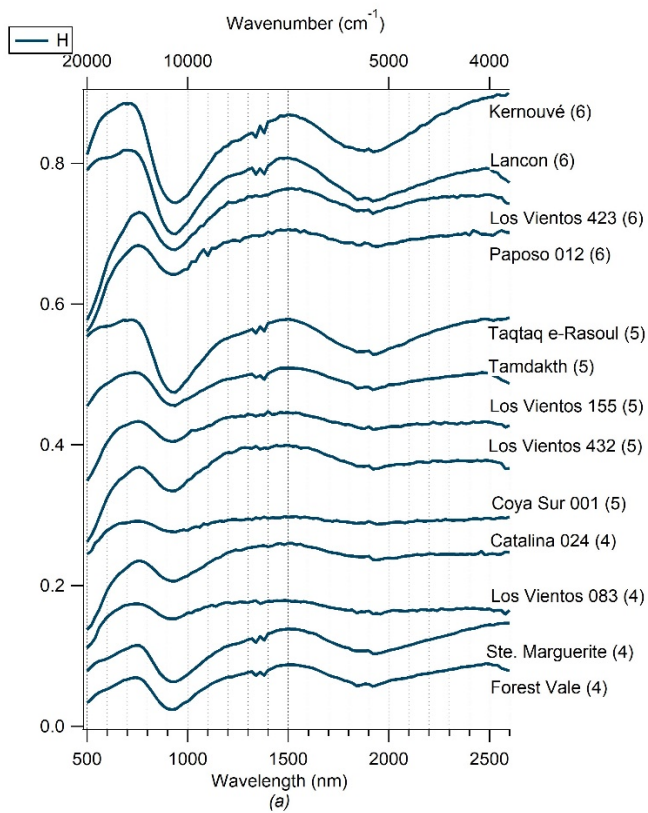
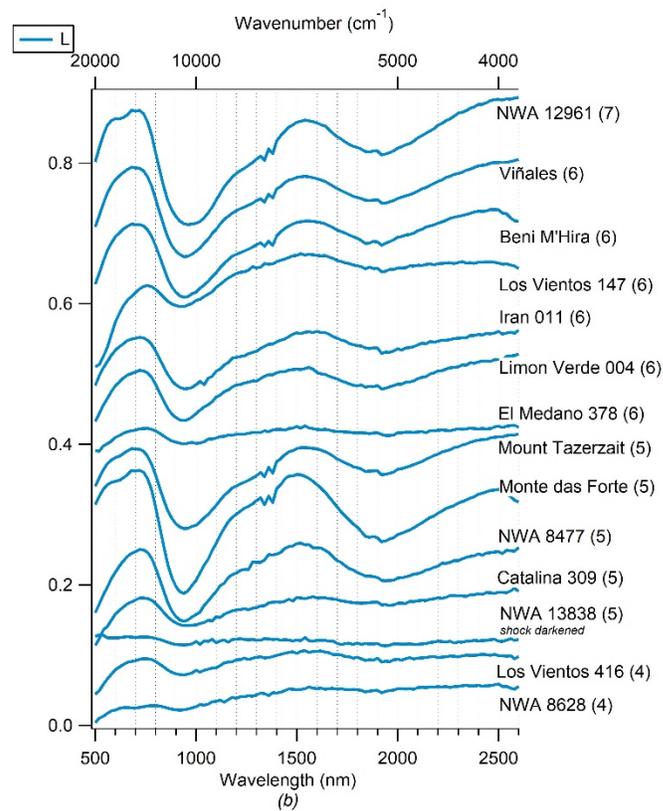
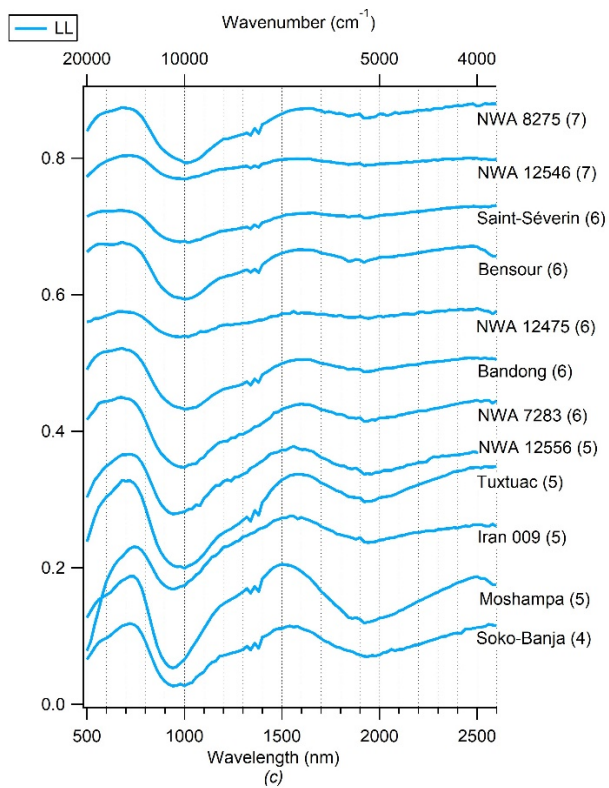

*Figure 2: Reflectance spectra of the 38 EOCs measured in the present work in the 500 – 2600 nm wavelength range. The spectra are shifted along the y-axis for better visibility and sorted by metamorphic grade within each class (H, H(L), L, LL(L) and LL). The metamorphic grade increases from bottom to top as indicated by the Petrographic type (PT) values (Bonal et al., 2016) given in parenthesis behind each sample name. The classification used in this graph is the same as determined in the present work (see Table 1)*

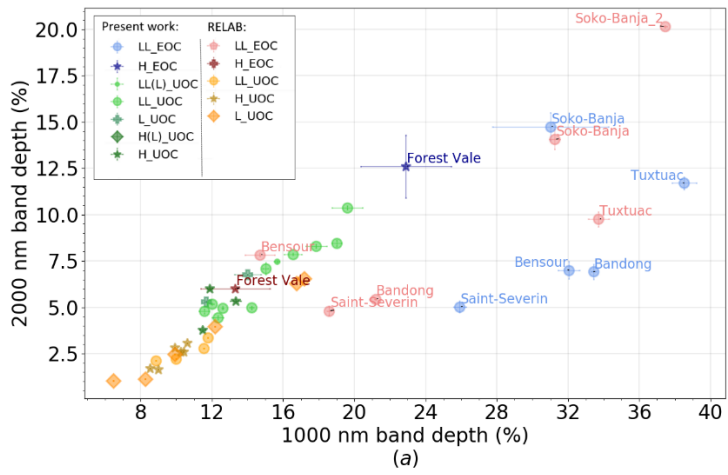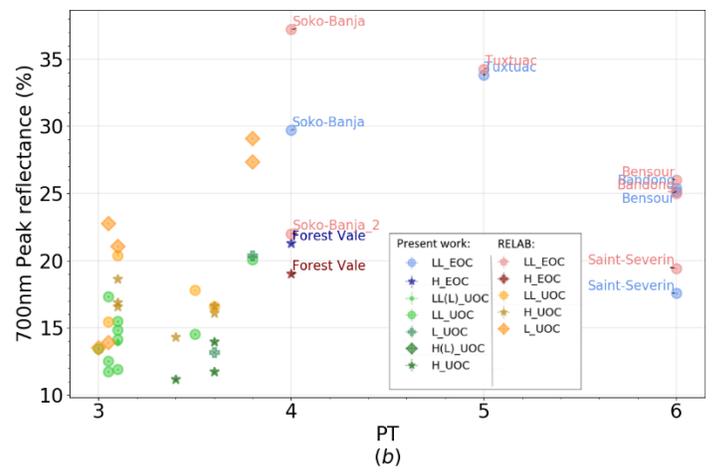

Figure 3: Comparison of the spectral features of the UOCs and EOCs measured in the present work with those already present in the RELAB database. (a) 2000 nm band depth (%) over the 1000 nm band depth (%) (b) 700 nm peak reflectance (%) over the PT.

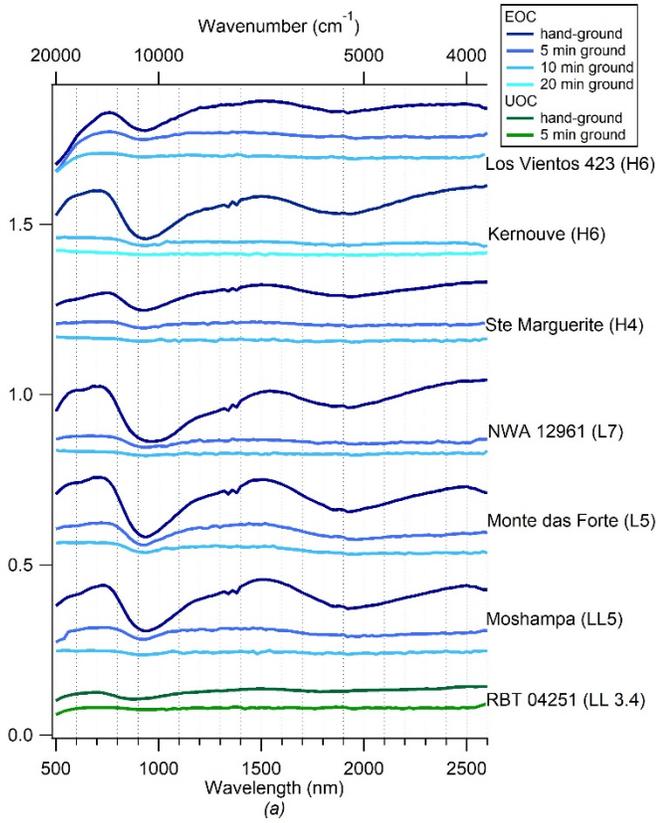
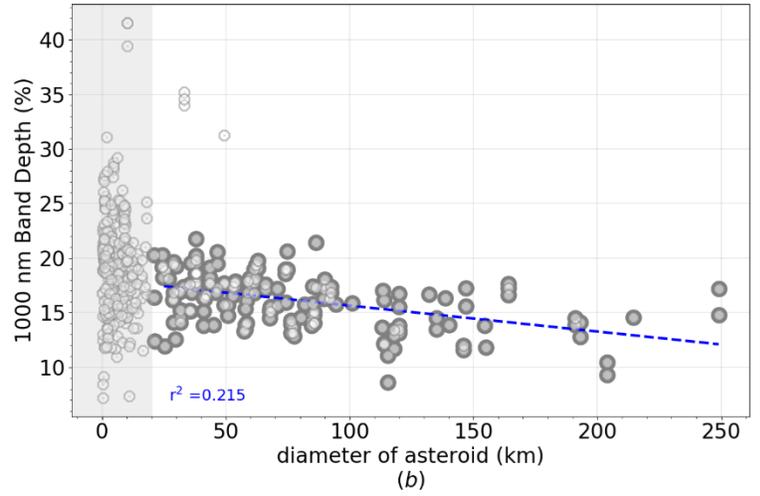

*Figure 4: (a) Reflectance spectra of EOCs Kernouvé, Ste Marguerite, NWA 12961, Monte das Forte, Moshampa and Los Vientos 423 as well as UOC RBT 04251 at different grain sizes. For each sample a spectrum is shown after being hand ground and ground for 5 min, 10 min and 20 min in a ball mill. The spectra are shifted along the y-axis for better visibility and samples were sorted by increasing metamorphic grade from bottom to top. (b) the 1000 nm band depth of the S-type asteroids over their diameter. Only asteroids highlighted in dark gray were used for the fit.*

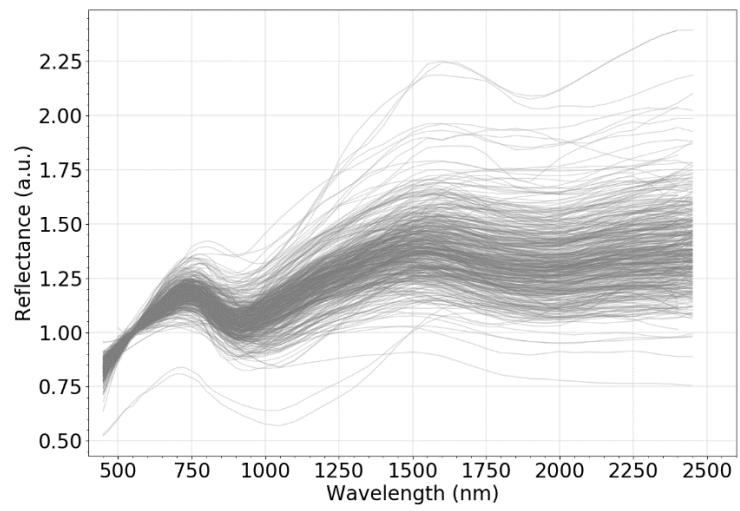 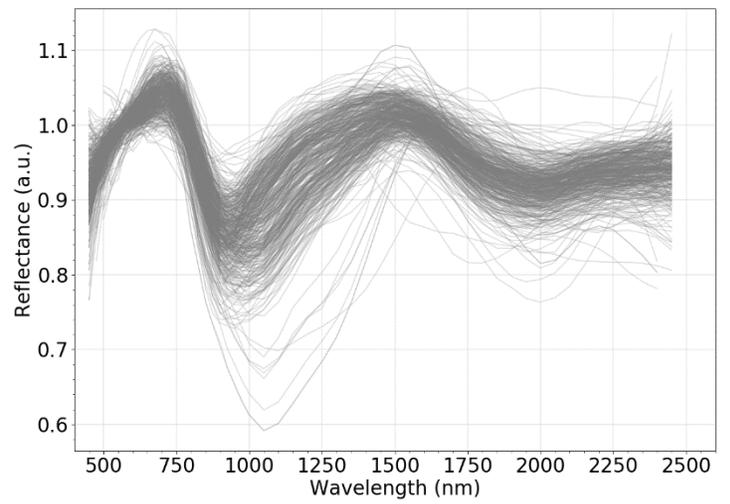

*Figure 5: (a) Reflectance spectra of the 466 raw S-type asteroid spectra considered in the present work. (b) Reflectance spectra of the 466 de-space weathered S-type asteroid spectra considered in the present work.*

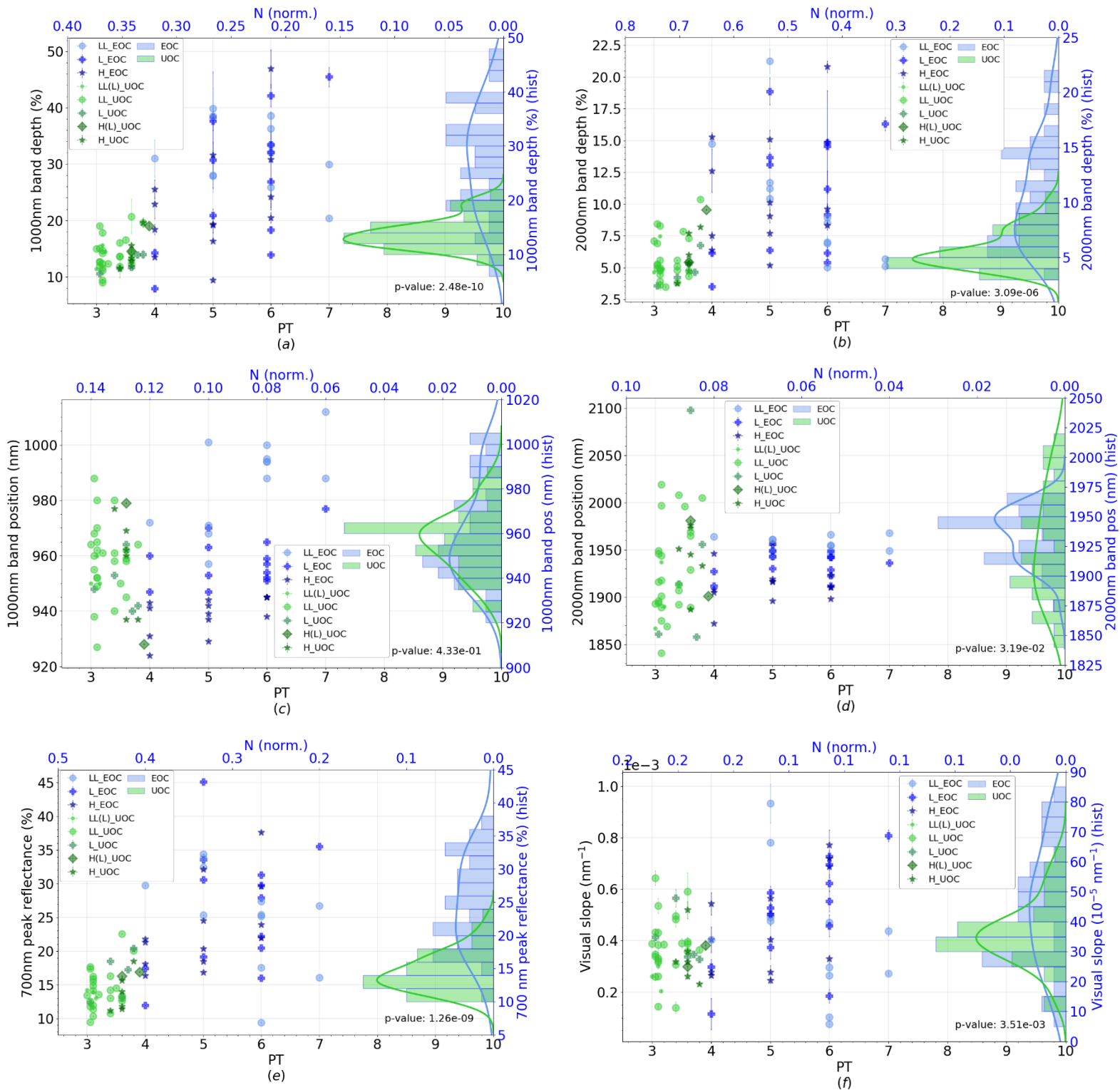

Figure 6: Comparison of the spectral features determined for UOCs and EOCs with their petrologic type (PT). Shown are: (a): The 1000 nm band depth over the PT, (b): The 2000 nm band depth over the PT, (c): 1000 nm band position over the PT, (d): 2000 nm band position over the PT, (e): 700 nm peak reflectance over the PT and (f): Visual slope over the PT. For each plot the y-axis is depicted as a histogram on a second y-axis to illustrate the separation between UOCs and EOCs. Each histogram is plotted with its gaussian kernel density estimate (kde). A 2D Kolmogorov Smirnof Test (2D KS-test) was performed for each spectral feature and the p-value is given as a measure of the distance between the two histograms in each plot.

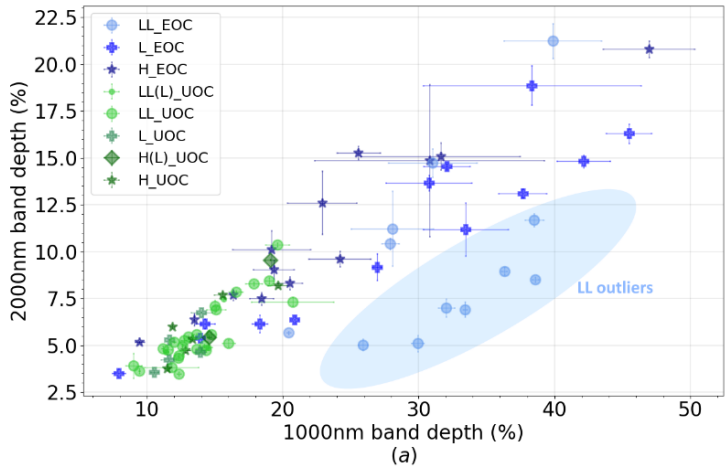 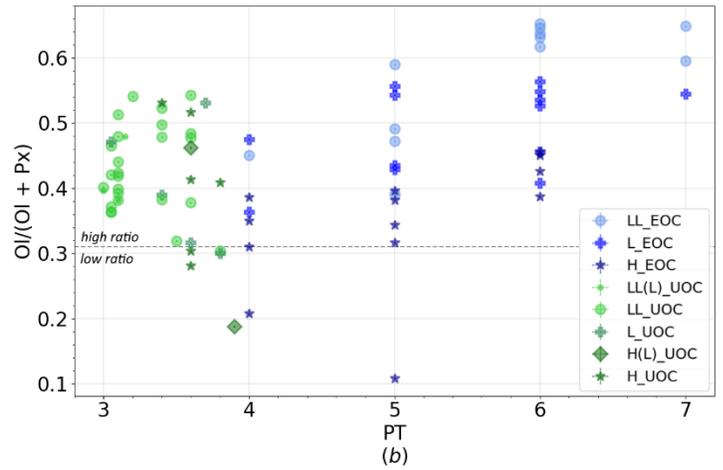

*Figure 7: (a) 2000nm band depth over the 1000 nm band depth of UOCs and EOCs. (b) Ol/(Ol + Px) ratio over the PT for EOCs and UOCs.*

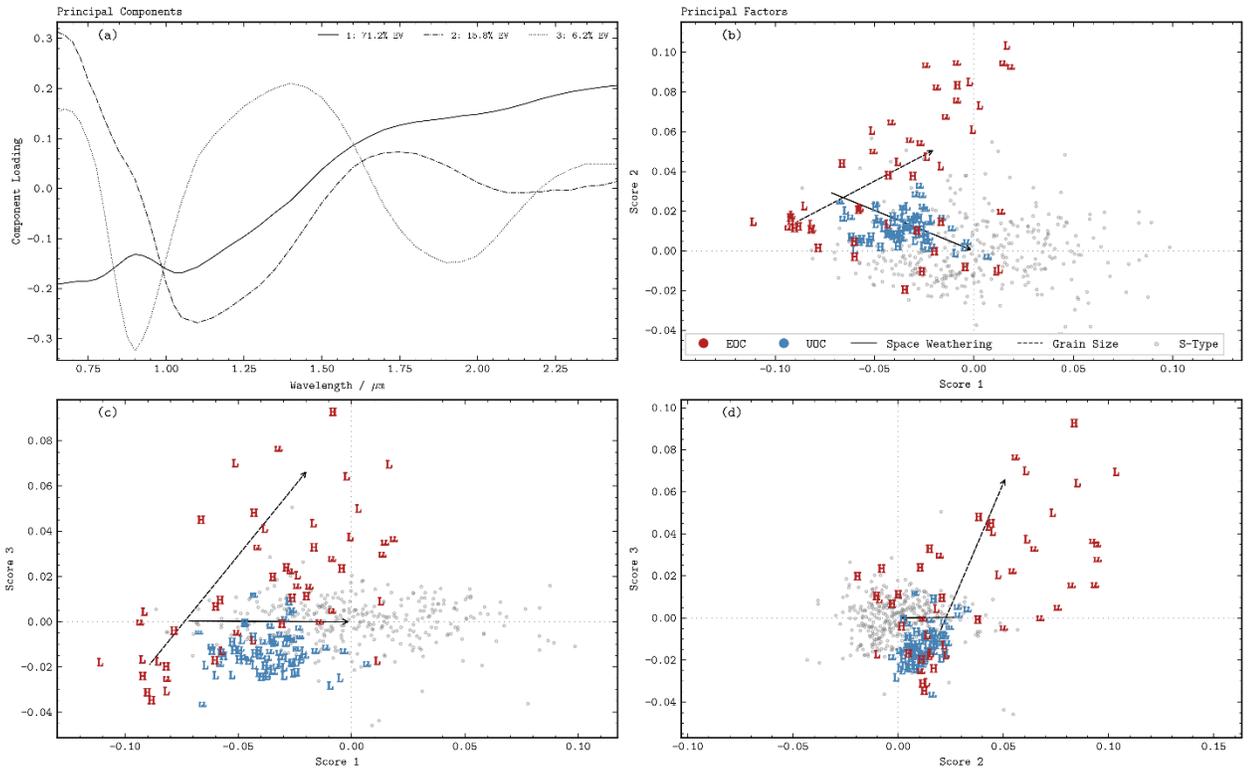

*Figure 8: Principal Components Analysis of the S-type and OC spectra. (a) The first three principal components computed from the S-type spectra. The labels give the percentage of explained variance (EV) for each component. (b), (c) and (d) show the principal scores of the S-type spectra (gray circles), EOCs (red) and UOCs (blue). The symbols give the mineralogic group of each spectrum. The solid black and dashed black arrows give the directions of increased space weathering and grain size respectively, as outlined in the text.*

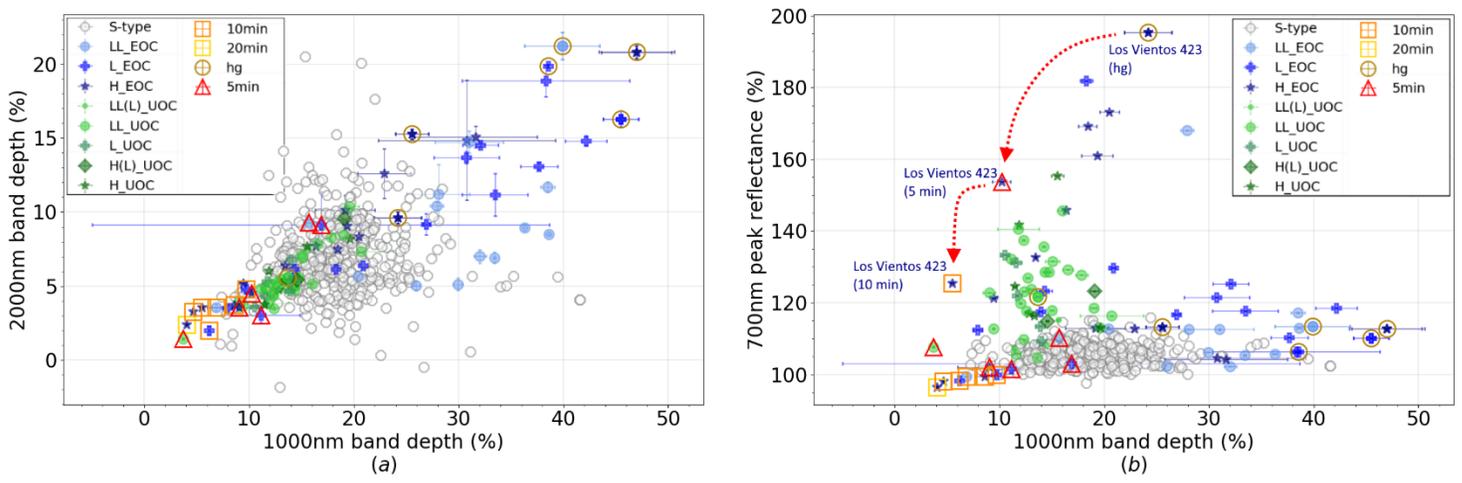

*Figure 9: Comparison of the de-space weathered S-type asteroids with EOCs and UOCs. (a) 2000 nm band depth over the 1000 nm band depth. (b) 700 nm peak relative reflectance (of spectra normalized to 550 nm) over the 1000 nm band depth. Samples that were measured at different grainsizes are outlined as follows: hand-ground (hg) by brown circles, 5 min ground in a ball mill by red triangles, 10 min ground in a ball mill by orange squares and 20 min ground in a ball mill by yellow squares.*

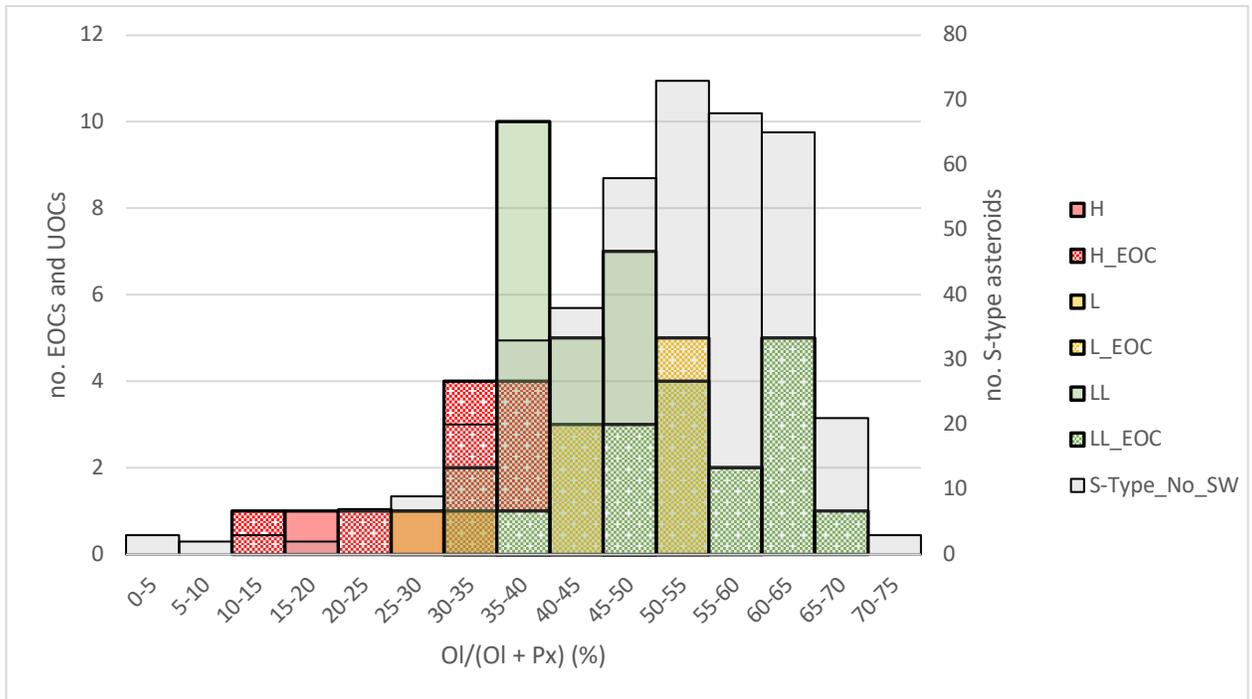

*Figure 10: Histogram of the Ol/(Ol + Px) ratio of S-type asteroids (on the right axis) as well as UOCs and ECOs (on the left axis). UOCs are plotted in solid colors, EOCs in pattern fill. S-type asteroids are plotted in solid grey color. Since the number of S-type spectra greatly exceeds those of EOCs and UOCs the histograms are plotted on two axes.*